\documentclass[10pt,superscriptaddress,onecolumn,showpacs,showkeys,aps,prl,preprint]{revtex4-1}

\usepackage{graphicx}% Include figure files
\usepackage{dcolumn}% Align table columns on decimal point
\usepackage{subfigure}
\usepackage{bm}
\usepackage{ifpdf,braket}
\usepackage{epstopdf}

\ifpdf
\usepackage[pdftex,
        colorlinks=true,
        pdftitle={Excited vortex states in superconductor},
        pdfauthor={Mauro M. Doria},
        baseurl={http://www.if.ufrj.br/~mmd},
        pdftoolbar=true,
        pdfmenubar=false,
        pdfpagemode=UseOutlines,
        pdfstartview=FitH,
        bookmarksopen=false
        ]{hyperref}
\fi

\begin{document}

\title{Paramagnetic excited vortex states in superconductors}

\author{Rodolpho Ribeiro Gomes}
\affiliation{Departement Fysica, Universiteit Antwerpen,
Groenenborgerlaan 171, B-2020 Antwerpen, Belgium}
\affiliation{Instituto de F\'{\i}sica, Universidade Federal do Rio de Janeiro, 21941-972 Rio de Janeiro, Brazil}%
\email{mauromdoria@gmail.com}
\author{Mauro M. Doria}
%\thanks{CNPq support from funding 23079.014992/2015-39.}
\affiliation{Dipartimento di Fisica, Universit\`a di Camerino, I-62032 Camerino, Italy}
\affiliation{Instituto de F\'{\i}sica, Universidade Federal do Rio de Janeiro, 21941-972 Rio de Janeiro, Brazil}%
\email{mmd@if.ufrj.br}
\author{Antonio R. de C. Romaguera}%
\affiliation{Departamento de F\'{\i}sica, Universidade Federal Rural de Pernambuco,52171-900 Recife, Pernambuco, Brazil}
%%\date{\today}

\begin{abstract}
We consider excited vortex states, which are vortex states left inside a superconductor once the external applied magnetic field is switched off and whose energy is lower than of the normal state. We show that this state is paramagnetic and develop here a general method to obtain its Gibbs free energy through conformal mapping. The solution for any number of vortices in any cross section geometry can be read off from the Schwarz - Christoffel mapping. The method is based on the first order equations used by A. Abrikosov to discover vortices.
\end{abstract}

%74.20.De	Phenomenological theories (two-fluid, Ginzburg-Landau, etc.)
%74.25.Ha	Magnetic properties including vortex structures and related phenomena (for vortices, magnetic bubbles, and magnetic domain structure, see 75.70.Kw)
%74.25.Uv	Vortex phases (includes vortex lattices, vortex liquids, and vortex glasses)
%74.25.Wx	Vortex pinning (includes mechanisms and flux creep)
%75.20.-g	Diamagnetism, paramagnetism, and superparamagnetism

\maketitle
\section{introduction}
Individual vortices in type II superconductors were first seen by U. Essmann and H. Trauble through the Bitter decoration technique~\cite{essmann67,essmann68}. Since then several other techniques have been developed for this purpose~\cite{bending99},
such as scanning SQUID microscopy~\cite{vu93}, high-resolution magneto-optical imaging~\cite{goa01,olsen04}, muon spin rotation ($\mu$SR)~\cite{sonier97,niedermayer99}, scanning tunneling microscopy~\cite{straver08,suderow14}, and  magnetic force microscopy~\cite{straver08}.
These advancements in the visualization of individual vortices opens the gate to investigate new properties~\cite{cren09,cren11}, such as those of the {\it excited vortex state} (EVS).
The EVS brings a new paradigm to the study of the transient dynamics of vortices in superconductors with boundaries, a subject of current interest due to the onset of instabilities~\cite{lukyanchuk15}.\\

A type II superconductor in presence of an external applied magnetic field contains vortices in its interior whose density is fixed by the external applied magnetic field. Once the applied field is switched off this state becomes unstable and vortices must leave the superconductor. However their exit can be hindered by microscopic inhomogeneities which pin them inside the superconductor. Here we make an important distinction concerning this left vortex state according on how its energy compares with that of the normal state. Although the left vortex state is always unstable, only in case its energy is lower than that of the normal state we call it EVS.
Consider the left vortex state immediately after the applied field is switched off. Vortices have topologically stability but only inside the superconducting state. Thus as long as the superconducting state exists they cannot be created or destroyed and therefore can only enter or exit through the boundary. However if the superconducting state ceases to exist vortices vanish all together inside the material.
Therefore the  obtainment of the Gibbs free energy difference between the superconducting state, which contains the left vortex state inside, and the normal state must be done to determine whether the superconducting state still exists.
In case this energy difference is positive one expects the collapse of the superconducting state and the onset of the normal state, possibly because of non-linear thermal effects that will develop and pinning will not be able to uphold this transition.
In case this energy difference is negative the EVS exists and can remain inside the superconductor provided that pinning impedes the motion of vortices towards the superconductor boundary.\\

In the London limit the energy of the vortex state is clearly positive since it is the sum of all two-vortex repulsive interactions. This is equivalent to say that the energy is the sum of the vortex lines self energies plus the sum over the  repulsive interaction between different vortices. Once again it becomes clear that vortices inside the superconductor can only be sustained by the external pressure exerted by the applied field and once this pressure ceases to exist they go away leaving behind the state of no vortices. However the study of the EVS cannot be done in this context of the London theory since this theory does not describe the condensate energy. Within the London theory
it is not possible to compare the energy of the vortex state with the normal state. Therefore the EVS must be studied in the context of the Ginzburg-Landau (GL) theory.\\

The EVS is intrinsically paramagnetic~\cite{thompson95,fabio04,balicas13}, which means that its magnetization points in the same direction of the switched off applied field. A simple way to understand this property is to analyse the two types of currents in a superconductor. In the Meissner phase there are only shielding currents at the boundary, which are diamagnetic, namely, they create an equal and opposite field to the applied one in its interior resulting into a net null field.
However the current around the vortex core is opposite to the shielding current, a property that explains why the increase in the number of vortices weakens the diamagnetic response set by the shielding currents. This increase reaches the critical point where the magnetization vanishes and then  the upper critical field is reached. Consider, for the sake of the argument, that after the sudden switch off the applied field the shielding currents immediate  disappear. This must occur in some moment otherwise there will be a left field  of opposite direction to the applied one. The vortices caught in this sudden transformation either remain pinned or start their move towards the boundary.
In case the Gibbs energy difference is positive the whole superconducting state will move towards its collapse, but if negative there will be a EVS left inside, which corresponds to a vortex state without the shielding currents and therefore with a purely paramagnetic response.
There is no Lorentz force to push vortices at the boundary since the shielding currents are absent. Vortices have no impediments at the boundary,  there is no Bean-Livingstone barrier in zero applied field~\cite{deo99,hernandez02}.\\

In this paper we propose a general method to calculate the Gibbs free energy difference between the left vortex and the normal states  valid near the superconducting to normal transition, where the order parameter is small.
This method is general since it applies to any vortex configuration in any geometry of the cross section area.
The method applies to very long superconducting cylinders such that the vortex lines are parallel to each other and oriented along its major axis.
As shown here, the order parameter that describes the left vortex state can be obtained from the well-known mathematical problem of conformal mapping. Remarkably the order parameter is just an analytical function with constant modulus at the boundary of a given cross section geometry.
In power of the order parameter we obtain the Gibbs energy difference, the local  magnetic field and currents inside the superconductor. Besides we also obtain other interesting features, such as the magnetic field at the center of the vortex core, and also the paramagnetic magnetization. It is well-known that at low vortex density the magnetic field inside the vortex core is twice the value of the lowest critical field~\cite{brandt95}. For higher densities a vortex lattice sets in and the magnetic field inside the vortex core varies according to the parameter $\kappa$ (the ratio between the penetration and the coherence lenghts) and the vortex density~\cite{brandt97}. Here we report the surprising result that as vortices move towards the boundary the magnetic field at their cores, and also the paramagnetic magnetization, change according to their positions. The magnetization vanishes when vortices reach the boundary.

The present approach stems directly from A. Abrikosov's seminal work~\cite{abrikosov57} that led to the discovery of vortices. There he found two identities that we refer here as the first order equations. They were later rediscovered by E. Bogomolny~\cite{bogomolny76} in the context of string theory and found to solve exactly the second order variational equations, that stem from the free energy, for a particular value of the coupling ($\kappa=1/\sqrt{2}$). These first order equations are able to determine the order parameter and the local magnetic field.
A. Abrikosov used them just in case of bulk superconductor, which means a superconductor with no boundary to a non superconducting region. Therefore he took the assumption of a lattice such that periodic boundary conditions apply. Obviously the bulk is an idealized system that simplifies the theory but hinders important boundary effects. Here we essentially extend this very same treatment to the case of a superconductor with a boundary.
Interestingly in case of the bulk, and of no applied external field, the first order equations predict no vortex solution. The only possible solution is that of a constant order parameter. However, as shown here, the existence of a boundary changes dramatically the scenario and vortex solutions become possible even without the presence of an applied field. Thus the profound connection between the mathematical theory of conformal mapping and the vortex solutions relies on the existence of boundaries.
The theory of conformal mapping was developed in the nineteenth century.
The Riemann mapping theorem of 1851 states that  any simply connected region in the complex plane can be conformally mapped onto any other, provided that neither is the entire plane. The mapping useful to us is the one that takes any finite geometry into the disk~\cite{boutsianis08} which is essentially a variant of the Schwarz – Christoffel conformal transformation~\cite{driscoll02} of the upper half-plane onto the interior of a simple polygon. The Schwarz – Christoffel mapping~\cite{trefethen80} is used in potential theory and among its applications are minimal surfaces and fluid dynamics.\\

Although we are mainly interested here in the EVS problem, it must be stressed that  the present method also applies in presence of an  applied external field. The only condition imposed is to be near to the superconducting to normal transition, i.e., near to the upper critical field line. The zero field case just corresponds to  particular case near to the the critical temperature, T$_c$.
However in the general case the connection between the order parameter solution and the conformal mapping only holds in case of a circular cross section.
Recently a new method to solve the linearized GL problem for
mesoscopic superconductors was proposed by means of conformal mapping~\cite{pereira13}. Thus our approach is distinct from this one since we are also able to obtain the local magnetic field from the order parameter whereas the above method is not.

We find useful to summarize this paper as follows. We start  describing the GL theory in section \ref{secondorder} and in section \ref{dualview} introduce the key element to our approach, which is the decomposition of the kinetic energy into a sum of three terms, namely, the ground state condition plus the magnetic interaction plus the boundary term. This decomposition is directly related to the first order  equations. In subsection \ref{boundary} we discuss the role played by the boundary conditions in the context of the first order equations and obtain expressions for the magnetization and the Gibbs free energy.
Dimensionless units are introduced in section \ref{dimensionless}, which are useful in the treatment of two examples of conformal mapping considered here. There the limit of weak order parameter and small field is shown to help in solving the first order equations.
Until this point the proposed method is very general and applies in case of an applied field as well. Only in section \ref{conformal} the external field is switched off and the connection between the first order  equations and the conformal mapping established.
In order to show the power of the present method we study two examples in section \ref{examples}, whose solutions are obtained analytically. Both have the cross section geometry of a disk, the first one with a vorticity $L$ at the center, subsection \ref{center}, and the other with vorticity one in any position is discussed in subsection \ref{offcenter}. We reach conclusions in section \ref{conclusion} and in the appendices \ref{appendix_a} and \ref{appendix_b} details of our analytical calculations are given.

%%%%%%% figures%%%%%%%%%%%%%%%%%%%%%%%%%%%%%%%%%%%%%%%%%%%%
%\begin{figure}[htbp]
\section{The Gibbs free energy of a long superconductor and the variational second order  equations}\label{secondorder}
Effects of the top and the bottom of a very long superconductor are neglected  such that symmetry along the major axis is assumed. Any cross sectional cut at a given $x_3$ plane reveals the same area $\Sigma$ and the same physical properties. The external constant magnetic field is oriented along the major axis, $\vec H = H_a\hat x_3$. Hence the  order parameter is only expressed by coordinates in this plane, $\psi(x_1,x_2)$, and the only one non-zero component of the local magnetic field is perpendicular to this plane, $h_{3}(x_1,x_2)=\partial_{1}A_{2}(x_1,x_2)-\partial_{2}A_{1}(x_1,x_2)$.
The difference between the superconducting and normal Gibbs free energy densities, defined as $\Delta\mathcal{G}$, is given by,
\begin{eqnarray} \label{gibbs0}
 \Delta \mathcal{G} &\equiv & \mathcal{G}_s -\mathcal{G}_n =  \int_{\Sigma}\frac{d^{2}x}{\Sigma}\left \{\alpha (T) |\psi|^2 + \frac{\beta}{2}|\psi|^4 \right .  \nonumber \\
&+& \left . \frac{|\vec{D}\psi |^2}{2m} +\frac{\left (h_{3}- H_{a}\right )^2}{8 \pi}\right \}. \label{GibbsFEdensity}
\end{eqnarray}
The normal state, $\mathcal{G}_s=\mathcal{G}_n $, is reached for  $\psi=0$ and $h_3=H_{a}$.
The well-known GL parameters have the properties that $\beta >0$ and that,
\begin{eqnarray}
\alpha(T)=\alpha_{0}\left( \frac{T}{T_c}-1\right) \Rightarrow \alpha(T) \, \mbox{is} \left\{ \, \begin{array}{c} <0 \,\mbox{ for }\, T  <  T_c  \nonumber\\
>0 \mbox{ for } \,T  >  T_c \, \end{array} ,\right.\nonumber
\end{eqnarray}
where $\alpha_0 $ is a positive constant.
The vector notation is two-dimensional, such as $\vec D = D_1 \hat x_1 + D_2 \hat x_2$ with
$D_j\equiv \left (\hbar/i \right)\partial_j\;-\; q\,A_j(x_1,x_2)/c$, $j=1,2$.
The well-known second order equations are obtained by doing variations with respect to the fields, namely, $\delta \vec A$ and $\delta \psi^{*}$, and lead to the non-linear GL equation,
\begin{eqnarray}
\frac{\vec{D}^2\psi}{2m} + \alpha \psi + \beta|\psi|^2 \psi=0, \label{gl-eq}
\end{eqnarray}
and to Amp\`ere's law,
\begin{eqnarray}
\vec \nabla \times \vec h = \frac{4\pi}{c} \vec J, \label{am-eq}
\end{eqnarray}
where the current density is given by,
\begin{eqnarray}
\vec{J}=\frac{q}{2m}\left (\psi^*\vec{D}\psi + c.c. \right). \label{curr0}
\end{eqnarray}
Boundary conditions must be added to find the physical solutions.
They correspond to no current flowing out of the superconductor, as vacuum is assumed outside, and the local field inside must be equal to the applied field outside. Let $\Upsilon$ be the perimeter of area $\Sigma$, thus at the boundary it must hold that,
\begin{eqnarray}
&& \hat n \cdot \vec{J}\vert_{\vec x \,\mbox{at}\,\Upsilon}=0, \, \mbox{and}, \label{bc-j}\\
&& h_3\vert_{\vec x \,\mbox{at}\,\Upsilon} = H_a, \label{bc-h3}
\end{eqnarray}
where $\hat n$ is perpendicular to the boundary.
Since the current is given by Eq.(\ref{curr0}) the following condition on the derivative of the order parameter,
\begin{eqnarray}
\hat n \cdot \vec{D}\psi\vert_{\vec x \,\mbox{at}\,\Upsilon}=0, \label{bc-dpsi}
\end{eqnarray}
is enough to guarantee the condition of Eq.(\ref{bc-j}).

\section{Dual view of the kinetic energy and the first order equations}\label{dualview}
The kinetic energy density admits a dual formulation due to the  mathematical identity proven in the appendix, given by Eq.(\ref{apwl}),
\begin{eqnarray}
&&\frac{|\vec{D}\psi|^2}{2m}=\frac{|{D}_{+}\psi|^2}{2m}+\frac{\hbar q}{2 m c}h_3|\psi|^2+\frac{\hbar}{2 q}\left (\partial_1 J_2 - \partial_2J_1 \right ),\quad
\label{math-iden}
\end{eqnarray}
where  ${D}_{+}\equiv {D}_{1}+i\,{D}_{2}$ and the current is given by Eq.(\ref{curr0}).
This decomposition of the kinetic energy as a sum of three terms is exact, and its derivation~\cite{alfredo13} is given in the appendix \ref{appendix_a}. Therefore the kinetic energy density,
\begin{eqnarray}
F_k= \int_{\Sigma} \frac{d^2x}{\Sigma}\,\frac{|\vec{D}\psi|^2}{2m},
\end{eqnarray}
is also given by,
\begin{eqnarray}
F_k= \int_{\Sigma} \frac{d^2x}{\Sigma}\,\left( \frac{|{D}_{+}\psi|^2}{2m}+\frac{\hbar q}{2 m c}h_3|\psi|^2  \right)+ \frac{\hbar}{2 q} \frac{1}{\Sigma}\oint_{\Upsilon} d\vec l \cdot \vec J.\nonumber \\ \label{dualkin}
\end{eqnarray}
In this dual view there is a superficial (perimetric) contribution due to the current. For a bulk superconductor, where $\Sigma \rightarrow \infty$, the superficial current vanishes either because the currents are localized within some region in the bulk away from the boundary or because of periodic boundary conditions, the latter case being that one considered by Abrikosov~\cite{abrikosov57}.
However in case of a finite area, such as for a mesoscopic superconductor, the current at the boundary must be considered and does not vanish.

The most interesting property of this dual view of the kinetic energy is that the current also acquires a new formulation. The current can be simply obtained by the variation of the kinetic energy with respect to the vector potential,
\begin{eqnarray}
\delta F_{k}=-\frac{1}{c} \int_{\Sigma} \frac{d^{2}x}{\Sigma} \vec{J}\cdot\delta\vec{A},
\end{eqnarray}
and use of the dual formulation leads to
\begin{eqnarray}
&J_1&=\frac{q}{2m}[(D_+\psi)^*\psi+\psi^*(D_+\psi)]-\frac{\hbar\,q}{2m}\partial_2|\psi|^2, \label{curr1} \\
&J_2&=i\frac{q}{2m}[(D_+\psi)^*\psi-\psi^*(D_+\psi)]+\frac{\hbar\,q}{2m}\partial_1|\psi|^2. \label{curr2}
\end{eqnarray}
The superficial current term does not contribute since at the edge the variation of the vector potential vanishes, namely, $\delta \vec A \ =0$, for $\vec x$ at $\Upsilon$.

In this paper we seek the minimum of the free energy through solutions of the first order equations, given below,
\begin{eqnarray}
&& D_{+}\psi =0 \label{D+psi}, \, \mbox{and},\\
&& h_3=H'-2\pi\frac{\hbar q}{m c} |\psi|^2, \label{Hlocal}
\end{eqnarray}
instead of solutions of the second order equations, given by Eqs.(\ref{gl-eq}) and (\ref{am-eq}).
On Eq.(\ref{Hlocal}), $H'$ is a constant to be determined but whose interpretation is very clear. It is the local field at the vortex core since there $\psi=0$, and so, $h_3=H'$. In the absence of vortices the order parameter is constant and $H'=2\pi(\hbar q/m c) |\psi|^2$ since it must hold that $h_3=0$ everywhere. We show that the above equations provide an exact solution to Amp\`ere's law and an approximate solution to the non-linear GL equation, and so, provided an easy and efficient method to search for the minimum of the free energy.

Amp\`ere's law is exactly solved and to see it, just take the condition of Eq.(\ref{D+psi}) into the current, as given by Eqs.(\ref{curr1}) and (\ref{curr2}). The Amp\`ere's law, given by $\partial_2 h_3=4\pi J_1/c$  and $\partial_1 h_3=- 4 \pi J_2/c $,  becomes $ \partial_2 h_3 = -(4\pi\hbar q /2mc) \partial_2|\psi|^2 $ and $\partial_1 h_3 = -(4\pi\hbar q /2mc)\partial_1|\psi|^2$, respectively, since Eq.(\ref{Hlocal}) holds. Then one obtains that,
\begin{eqnarray} \label{curr-foe}
\vec J = \frac{\hbar q}{2 m} \hat x_3 \times \vec \partial |\psi|^2.
\end{eqnarray}
The surface term, contained in the dual formulation of the kinetic energy, is given by Eq.(\ref{math-iden}).
\begin{eqnarray}
\frac{\hbar}{q} (\partial_1 J_2 - \partial_2 J_1) =
+\frac{\hbar^2}{2 m}{\vec \partial}\,^2 |\psi|^2.
\end{eqnarray}
The mathematical identity of Eq.(\ref{math-iden}), becomes,
\begin{eqnarray} \label{fkin02}
\frac{|\vec{D}\psi|^2}{2m}=\left(H' - \frac{h q}{m c}|\psi|^2\right)\frac{\hbar q}{2 m c}|\psi|^2+\frac{\hbar^2}{4 m}\vec{\partial}\,^{2}|\psi|^2,
\end{eqnarray}
once assumed that the first order equations are satisfied.
The non-linear GL equation is approximately solved in the sense that its integrated version is exactly solved. This integrated version is obtained by multiplying the non-linear GL equation with  $\psi^*$, and next integrating over the entire area $\Sigma$ of the superconductor:
\begin{eqnarray}
\int_{\Sigma}\frac{d^2 x}{\Sigma}\left \{\psi^*\frac{\vec{D}^2\psi}{2m} + \alpha(T) |\psi|^2 + \beta|\psi|^4\right \}=0.
\end{eqnarray}
Next we transform this equation into an algebraic one whose usefulness is to fix the scale of the order parameter which has remained undefined when solving the scale invariant Eq.(\ref{D+psi}). The first term of the integrated equation summed with its complex conjugate and divided by $2$, can be expanded as follows,
\begin{eqnarray}
&& \frac{1}{2}\left[\psi^{*}\frac{\vec{D}^2\psi}{2\,m}+\left( \frac{\vec{D}^2\psi}{2\,m}\right)^{*}\psi\right]=-\frac{\hbar^{2}}{4\,m}{\vec \partial}\,^{2}|\psi|^{2}+\frac{|\vec{D}\psi|^{2}}{2\,m}\label{KExpansion}.\quad\nonumber\\
\end{eqnarray}
Inserting Eq.(\ref{KExpansion}) into the integrated equation, upon summing with its complex conjugate and dividing by $2$, one obtains that,
\begin{eqnarray}
\int_{\Sigma}\frac{d^2 x}{\Sigma} \left \{- \frac{\hbar^{2}}{4\,m}{\vec \partial}\,^{2}|\psi|^{2}+\frac{|\vec{D}\psi|^2}{2m}+\alpha(T) |\psi|^2 + \beta|\psi|^4 \right \}=0.\nonumber \\
\end{eqnarray}
The use of  Eq.(\ref{fkin02}) turns the integrated equation into the following one.
%\begin{eqnarray}\label{integrate0}
%&&\int_{\Sigma}\frac{d^2 x}{\Sigma}\left \{ \left (  \frac{H'}{2}\frac{\hbar\,q}{m\,c}+ \alpha(T) \right)|\psi|^2 \nonumber \\
%&-& \left [ \pi\left(\frac{\hbar\,q}{\,m\,c}\right)^2-\beta \right ]  |\psi|^4   \right \}=0
%\end{eqnarray}
\begin{eqnarray}\label{integrate0}
\int_{\Sigma}\frac{d^2 x}{\Sigma}\left \{ \left(  \frac{H'}{2}\frac{\hbar\,q}{m\,c}+\alpha \right)|\psi|^2-\left[ \pi\left(\frac{\hbar\,q}{\,m\,c}\right)^2-\beta \right]  |\psi|^4   \right \}=0.\nonumber\\
\end{eqnarray}
In summary Eq.(\ref{D+psi}), together with Eq.(\ref{integrate0}), fully determines the order parameter, $\psi$, and from Eq.(\ref{Hlocal}) one obtains the local magnetic field, $h_3$.

\subsection{The boundary conditions, the magnetization and the Gibbs free energy}\label{boundary}
There should be no current flowing out of the superconductor and  the magnetic field must be continuous at the boundary. Here we address the question on how to satisfy these boundary conditions in the context of the first order equations. The boundary conditions themselves are first order equations, as seen in Eqs.(\ref{bc-j}) and (\ref{bc-h3}), and so, their fulfilment is easily understood in the context of the second order equations. For instance the derivative of the order parameter normal to the surface must vanish, according to Eq.(\ref{bc-dpsi}), but this condition cannot be imposed on $\psi$,  obtained through Eq.(\ref{D+psi}) because this is a first order equation itself and  there is not enough free parameters in this solution.
However it is possible to satisfy Eq.(\ref{bc-j}) simply by the requirement that the density $|\psi|^2$ be  constant at the boundary, which introduces a new parameter, $c_0$, fixed by the non-linear integrated GL equation, Eq.(\ref{integrate0}).
\begin{eqnarray}
&& |\psi|^2=c_0^2 \;\; \mbox{for}\; \vec x \,\mbox{at}\;\Upsilon \label{bc-foe1}, \, \mbox{and},\\
&& H'=H_a+2\pi\frac{\hbar q}{m c} c_{0}^2. \label{bc-foe2}
\end{eqnarray}
Therefore the constant $H'$ is automatically fixed by $c_0$  according to Eq.(\ref{Hlocal}). The important point is that Eq.(\ref{bc-foe1}) is enough to guarantee that there is no current flowing outside the superconductor. This is  just a direct consequence of  Eq.(\ref{curr-foe}). As $|\psi|^2$ is constant along the border there is no gradient tangential to it and only perpendicular to it, namely, $\vec \partial |\psi|^2$ is normal to the surface, rendering $\vec J$ always tangent to the surface.

The magnetization $M_3$ is also directly obtained from the present formalism and easily shown to be paramagnetic in the absence of an applied external field. According to Eq.(\ref{Hlocal}),
\begin{eqnarray}\label{mag1}
B_3 \equiv \int_{\Sigma} \frac{d^2x}{\Sigma}\,h_{3}=  H'-4\pi\mu_B\int_{\Sigma}\frac{d^2 x}{\Sigma}|\psi|^2,
\end{eqnarray}
given in units of the Bohr's magneton, $\mu_B=\hbar q/2mc$.
Just take the thermodynamical relation $B_3=H_a+4\pi M_3$ and Eq.(\ref{bc-foe2}) to obtain that,
\begin{eqnarray}\label{mag}
M_3 =  \mu_B\int_{\Sigma}\frac{d^2 x}{\Sigma} \left ( c_{0}^2-|\psi|^2\right ),
\end{eqnarray}
which means that the integral has the dimension of inverse volume.
In case of no applied external field, the order parameter is maximum at the boundary, namely ($|\psi|^2 \le c_{0}^2$), and so the magnetization is paramagnetic, $M_3>0$.

Under the condition that the first order equations are satisfied the Gibbs free energy of Eq.(\ref{GibbsFEdensity}) becomes,
\begin{eqnarray}
\Delta \mathcal{G}&=& \int_{\Sigma}\frac{d^{2}x}{\Sigma}\left \{ \left(\alpha (T)+\frac{H_{a}}{2}\frac{\hbar\,q}{m\,c}\right ) |\psi|^2\right .  \nonumber \\
&+& \left .\frac{1}{2}\left [ \beta- \pi\left(\frac{\hbar\,q}{\,m\,c}\right)^2 \right ]|\psi|^4\right \} + \frac{\left (H'- H_{a}\right )^2}{8 \pi} \nonumber \\
&+& \frac{\hbar^2}{4m}\oint_{\Upsilon} \,\frac{dl}{\Sigma} \hat n \cdot \vec \partial |\psi|^2.  \label{GibbsFEdensity-foe}
\end{eqnarray}
Two direct contributions of the boundary to the free energy, which are not present in A. Abrikosov's treatment of the GL theory~\cite{abrikosov57}, are the field energy due to $H' \neq H_a$, according to Eq.(\ref{bc-foe2}), and the perimetrical contribution of the normal  gradient of $|\psi|^2$.

Near to the transition to the normal state the order parameter is weak, fact that allows for the expansion of the free energy in powers of $\psi$. From the present point of view this weakness also leads to the proposal of an iterative method to solve the first order equations.  Firstly one seeks a solution for $\psi$ in Eq.(\ref{D+psi}) under a known external field $H_0$ sufficiently near to the upper critical field line which sets the order parameter in the vicinity of the normal state transition.
Any solution of Eq.(\ref{D+psi}), multiplied by a constant is also  a solution and this constant is fixed by the integrated equation  (Eq.(\ref{integrate0})).
Eq. (\ref{D+psi}) can be rewritten as,
\begin{eqnarray}\label{D+psi4}
\left [ \left ( -i\frac{\partial}{\partial x_1}+\frac{\partial}{\partial x_2}\right)-\frac{2\pi}{\Phi_0}\left(A_1+iA_2 \right)\right ] \psi=0.
\end{eqnarray}
Let us introduce the complex notation, $z\equiv x_1+ix_2$, $\bar z\equiv x_1-ix_2$ into Eq.(\ref{D+psi4}), and consider a constant external field, $H_{a}$, such that $A_1=-H_{a} x_2/2$ and $A_2=H_{a} x_1/2$. In this case the above equation becomes,
\begin{eqnarray}\label{D+psi5}
 \frac{\partial\psi(z,\bar z)}{\partial \bar z}=-\left ( \frac{\pi H_a}{2\Phi_0}z \right) \psi(z,\bar z),
\end{eqnarray}
where $\partial / \partial z = 1/2 (\partial / \partial x_1 - i \partial /\partial x_2) $ and its complex conjugate is $\partial / \partial \bar z = 1/2 (\partial / \partial x_1 + i \partial /\partial x_2) $.
The solution of Eq.(\ref{D+psi5}) is promptly found to be,
\begin{eqnarray}\label{sol0}
\psi(z,\bar z)= f(z) e^{-\left(\frac{\pi H_a}{2\Phi_0} \right)z\bar z},
\end{eqnarray}
where $f(z)$ is {\it any} function of $z$.
The local field is equal to $H_a$ plus a correction proportional to $|\psi|^2$ according to Eq.(\ref{Hlocal}). Very near to the normal state one expect that this correction is small such that it suffices to solve Eq.(\ref{D+psi}) and feed the solution in Eq.(\ref{Hlocal}) without any further recurrence, namely, a returns to Eq.(\ref{D+psi}) with a corrected  $\vec A$ associated to the local field $h_3$.
Thus the quest for a solution $\psi(z,\bar z)$ in a given geometry with cross sectional area $\Sigma$ and boundary $\Upsilon$ is reduced to the search of an analytical function $f(z)$ that will render $|\psi(z, \bar z)|^2$  constant at the boundary $\Upsilon$.
Assuming that in the boundary $|\psi|^2=c_0^2$ this means that the analytical function $f(z)$ must satisfy  the condition $|f(z)|=c_0 \exp{\left(\frac{\pi H_a}{2\Phi_0}|z|^2\right)}$, where $z$ belongs to the boundary.
For a circular disk, $|z|$ is automatically constant at the boundary, but in case of an arbitrary geometry this characteristic is no longer satisfied.
In this paper we will only consider the case that there is no external applied field, $H_a=0$.

\section{The dimensionless units} \label{dimensionless}     % Enter section title between curly braces
At this point we find useful to switch to dimensionless units and rewrite all previous expressions in this fashion. As well known the GL theory has only one coupling constant, $\kappa = \lambda/\xi$,  the ratio between the London penetration length, $\lambda = \left (m c^2/4 \pi q^2 \psi_{0}^{2} \right)^{1/2}$,  and the coherence length, $\xi = \left ( \hbar^{2}/2m|\alpha| \right )^{1/2}$, respectively, where $\psi_0 = \left (|\alpha|/\beta \right )^{1/2}$. This renders this ratio temperature independent, $\kappa = \left ( \beta/2\pi \right )^{1/2}mc/\hbar q$ and only dependent on material parameters. Let us refer just in this paragraph to the dimensionless units by the prime notation. For instance distance is measured in terms of the coherence length such that $\vec{x}=\xi\vec{x}'$. The magnetic field is expressed in units of the upper critical field, $\vec h = H_{c2}\vec h'$, $H_{c2}=\sqrt{2}\kappa H_c=\Phi_0/2\pi\xi^2$, $\Phi_0=hc/q$, where $H_c = \Phi_0/2\pi\sqrt{2}\lambda\xi$ is the thermodynamical field. Then the dimensionless vector potential is given by $\vec{A}=H_{c2} \xi \vec{A}'$, and the covariant derivative becomes  $D_{j}= D'_{j}/\xi$, $D'_{j}\equiv\frac{1}{i}\partial'_{j}-A'_{j}$ since $\partial_{j}=\partial'_{j}/\xi$. Lastly the dimensionless order parameter is obtained from $\psi=\psi_0\psi'$. The order parameter value at the boundary is also defined dimensionless, $c_0=\psi_0 c_0'$, and finally the dimensionless magnetization is $M_3=M_3'H_{c2}/8\pi \kappa^2$ since $\psi_0^2\mu_B = H_{c2}/8\pi \kappa^2$.

Hereafter we drop the prime notation in all quantities, which means that they are all expressed in dimensionless units. The first order equations become,
\begin{eqnarray}
&& D_{+}\psi =0 \label{D+psi2}, \, \mbox{and},\\
&& h_3=H'-\frac{1}{2\kappa^2} |\psi|^2. \label{Hlocal2}
\end{eqnarray}
Once the order parameter and the local magnetic field satisfy the first order equations, the kinetic energy of Eq.(\ref{math-iden})  becomes,
\begin{eqnarray} \label{fkin2}
|\vec{D}\psi|^2=\left(H' - \frac{1}{\kappa^2}|\psi|^2\right)|\psi|^2+\frac{1}{2}\vec{\partial}\,^{2}|\psi|^2.
\end{eqnarray}
The integrated equation, Eq.(\ref{integrate0}), in reduced units becomes,
\begin{eqnarray}
&&\int_{\Sigma}\frac{d^2 x}{\Sigma}\left[\left(H'-1\right)|\psi|^2+\left(1-\frac{1}{2\kappa^2}\right)
|\psi|^4\right]=0.\nonumber\\
&&\label{ReducedIntegratedEquation2}
\end{eqnarray}
The Gibbs free energy difference of Eq.(\ref{gibbs0}) is in units of $\,H_{c}^{2}/4\pi\,=\,|\alpha|^2/\beta\,$, given by,
\begin{eqnarray}
\Delta\mathcal{G}&\equiv\frac{\mathcal{G}-\mathcal{G}_n}{|\alpha|^2/\beta}&= \int_{\Sigma}\frac{d^{2}x}{\Sigma}\left\{-|\psi|^2+
\frac{1}{2}|\psi|^4+|\vec{D}\psi|^2\right. \nonumber \\
&+&\left.\kappa^2 (h_{3}-H_{a})^2\right\}.
\end{eqnarray}
Similarly, Eq.(\ref{GibbsFEdensity-foe}), becomes
\begin{eqnarray}\label{GibbsFEdensity-foe2}
\Delta\mathcal{G}&=&\int_{\Sigma}\frac{d^{2}x}{\Sigma}\left\{-\left(1-H_a\right)|\psi|^2+\frac{1}{2}\left(1-
\frac{1}{2\kappa^2}\right)|\psi|^4\right.\nonumber \\
&+&\left.\frac{1}{2}{\vec\partial}\,^{2}|\psi|^2+\kappa^2(H'-H_{a})^2\right\}.\label{deltaGibbs2}
\end{eqnarray}

\section{The vortex excited state and conformal mapping}\label{conformal}
In the absence of an applied field outside the superconductor, $H_a=0$ , and with the help of Eq.(\ref{ReducedIntegratedEquation2}),  the Gibbs free energy density of Eq.(\ref{GibbsFEdensity-foe2}) becomes,
\begin{eqnarray}\label{GibbsFEdensity-foe3}
\Delta\mathcal{G} = \int_{\Sigma}\frac{d^{2}x}{\Sigma}\left\{-\frac{1}{2}\left( 1+H' \right)|\psi|^2+\frac{1}{2}{\vec\partial}\,^{2}|\psi|^2+\kappa^2 {H'}^2\right \}. \nonumber \\
\label{deltaGibbs3}
\end{eqnarray}
Then $H'$ follows directly from Eq.(\ref{bc-foe2}).
Near to the normal state  the order parameter is small enough that iteration of the first order equations is not necessary. This means that the local magnetic field is small enough that it can be dropped in Eq.(\ref{D+psi3}) and directly obtained from Eq.(\ref{Hlocal3}). In this case Eqs.(\ref{D+psi2}) and (\ref{Hlocal2}) are reduced to,
\begin{eqnarray}
&& \partial_{+}\psi =0 \label{D+psi3}, \, \mbox{and},\\
&& h_3= \frac{c_{0}^2- |\psi|^2}{2\kappa^2}. \label{Hlocal3}
\end{eqnarray}
The connection between vortex states and conformal mapping stems from Eq.(\ref{D+psi5}), which simply becomes,
\begin{eqnarray}
\frac{\partial \psi(z,\bar z)}{\partial \bar z}=0.
\end{eqnarray}

Expressing the order parameter as $\psi\equiv \psi_R+i\psi_I$ in Eq.(\ref{D+psi3}), one obtains that $\partial_1 \psi_R + \partial_2 \psi_I=0$ and $\partial_2 \psi_R - \partial_1 \psi_I=0$, which are the well-known Cauchy-Riemann conditions for analyticity.
We recall the so-called {\it Maximum Modulus Theorem} in Mathematics which says that for an analytical function $\psi(z)$ in a given region $\Sigma$, the maximum of $|\psi(z)|$ necessarily  falls in its boundary $\Upsilon$. This theorem is useful because from it we know  that the order parameter,  constant at the boundary $\Upsilon$, indeed reaches its maximum value there, and therefore the magnetization is necessarily paramagnetic.
The presence of vortices inside will only lead to the vanishing of the order parameter at points in its interior, $\psi=0$, and still the maximum of $|\psi|^2$ is at the boundary. This feature helps to establish fundamental differences between the finite boundary and the bulk superconductor, the latter understood as the case that only periodic solutions are sought. According to Liouville's theorem~\cite{fine97} any periodic analytical function $\psi$ with zeros must also diverge, from which one draws the conclusion that the only possible physical solution is the constant one. In other words it is not possible to find vortex solutions without the presence of an applied field in a unit cell with periodic boundary conditions. However they do exist for the long superconductor with a finite cross section. Thus we conclude from the above discussion that the present method is general and applies for any number of vortices in any cross sectional geometry.
As examples of our general method, we detail in this paper two particular cases of a disk, namely, a vortex with vorticity $L$ in its center and a vortex with vorticity one in any position inside the disk.
Let us define the integrals,
\begin{eqnarray}
I_2 &\equiv& \frac{1}{c_0^2}\int_{\Sigma}\frac{d^{2}x}{\Sigma}|\psi|^2, \label{inti2}\\
I_4 &\equiv& \frac{1}{c_0^4}\int_{\Sigma}\frac{d^{2}x}{\Sigma}|\psi|^4, \, \mbox{and}, \label{inti4} \\
I_{\Upsilon}&\equiv& \frac{1}{c_0^2}\int_{\Sigma}\frac{d^{2}x}{\Sigma}{\vec \partial}\,^{2}|\psi|^2 .\label{intis}
\end{eqnarray}
From the integrated equation, Eq.(\ref{ReducedIntegratedEquation2}) and the condition that the local magnetic field vanishes at the boundary, one obtains that,
\begin{eqnarray}
c_{0}^{2}&=&\frac{1-H'}{1-1/2\,\kappa^{2}}\frac{I_2}{I_4}, \, \mbox{and}\, , \\
H'&=&\frac{c_{0}^{2}}{2\kappa^2}.\label{Hlinha}
\end{eqnarray}
Solving these equations for $c_{0}$ and $H'$, we obtain that,
\begin{eqnarray}
c_{0}^{2}&=&\frac{2\kappa^2\,I_2/I_4}{2\kappa^{2}-1+I_2/I_4}, \label{c0PinA}\, \mbox{and}\, , \\
H'&=&\frac{I_2/I_4}{2\kappa^{2}-1+I_2/I_4}.\label{H*PinA}
\end{eqnarray}
Notice that $I_2>I_4$ is indeed satisfied since the order parameter divided by $c_0$ is always smaller than one inside the disk, and so, at each point the second power, Eq.(\ref{inti2}), is larger than the fourth power, Eq.(\ref{inti4}). The fact that  $I_2>I_4$ and $\kappa>1/\sqrt{2}$, always render a solution for $c_0$.
The difference of the free energy density given at Eq.(\ref{deltaGibbs3})  can be expressed in terms of $c_{0}$,
\begin{eqnarray}
\Delta\mathcal{G} &=&-\frac{1}{2}(I_{2}-I_{\Upsilon})\,c_{0}^{2}-\frac{1}{4\kappa^2}(I_{2}-1)c_{0}^{4} ,\label{GFEDensc0}
\end{eqnarray}
or in terms of $H'$,
\begin{eqnarray}
\Delta\mathcal{G} &=&-\kappa^2(I_{2}-I_{\Upsilon})\,H'-\kappa^2(I_{2}-1)H'^{2}.\label{GFEDensH*}
\end{eqnarray}
The magnetization given by Eq.(\ref{mag}) becomes $M_3=2\kappa^2H'-I_2c_0^2$, and so, equal to,
\begin{eqnarray}
M_3=\frac{2\kappa^{2}\left( 1-I_2 \right)I_2/I_4}{2\kappa^{2}-1+I_2/I_4}. \label{M3}
\end{eqnarray}
This interesting expression shows that the magnetization depends on the position of the vortices inside the superconductor, since $I_2$ and $I_4$ vary accordingly, as shown in the next example.
In case of no vortices, $I_2=I_4=1$, the magnetization vanishes.
We also include the general expression for the magnetic field at the center of the vortex, as $h_3(v)=H'$:
\begin{eqnarray}
h_3(v)=\frac{I_2/I_4}{2\kappa^{2}-1+I_2/I_4}.\label{h30}
\end{eqnarray}
We consider two particular examples of a thin long cylinder with radius $R$, namely,  a vortex $L$ at the center and a vortex $L=1$ at arbitrary position. Through them we see the general aspects of the present theory such as the importance of the boundary term which makes the Gibbs energy explicitly dependent on $R$.
Remarkably the cooper pair density at the boundary determines the local magnetic field at the vortex core. There $\psi(v)=0$ and so $h_3(v)=c_0^2/2\kappa^2$ according to Eq.(\ref{Hlocal2}) where $v$ refers to the center of the vortex.
%------------------------------------------------------------------------------
\section{Long cylinder with a circular cross section} \label{examples}
We find useful to apply our theory to Niobium~\cite{finnermore66,essmann72,das08}, one of the favored materials to study the characteristics of vortex matter in superconductors and also used to construct nano-engineered superconductors~\cite{bothner12}
All figures are expressed in reduced units and to retrieve the predicted values for  Niobium we take the parameter values reported in Ref.~\onlinecite{das08}, namely, $\kappa =2.1$, $\lambda(0)$ = 42 nm, $\xi(0)$ = 20 nm. In particular we choose the temperature of T = 7.7 K which is near to the critical temperature, T$_c$ = 9.3 K, such that an order parameter approach is valid. Thus for this temperature the local magnetic field, expressed in units of the upper critical field, must be multiplied by $H_{c2}(T=7.7 K)$= 780 Oe. Similarly the magnetization must be multiplied by $H_{c2}(T=7.7 K)/8\pi \kappa^2$= 7.0 Oe.

\subsection{Vorticity $L$ at the center}\label{center}
In this simple example we show that the EVS exists only in a special range of the radius and the vorticity. The search for the order parameter in a disk cross section of radius $R$ that satisfies  Eq.(\ref{D+psi3}), is reduced to find an analytical function that is constant at the perimeter of the disk. This is simply given by,
\begin{eqnarray}
\psi = c_0 \, \left (\frac{z}{R}\right )^L = c_0 \, \left (\frac{r}{R}\right )^L\,e^{iL\theta},\label{psiVortL}
\end{eqnarray}
where $c_0$ is the value of the order parameter at the boundary and $L$ is an integer since the order parameter is assumed to be single-valued, $\psi(\theta+2\pi)=\psi(\theta)$.
In polar coordinates the order parameter is expressed in terms of  $r \equiv \sqrt{z\bar z}$ and $\tan \theta \equiv  x_2/x_1$.
The fulfillment of $|\psi|=c_0$ at the circle guarantees the confinement of the current to the superconductor´s boundaries and $\psi=0$ at the center means that the solution has non-zero vorticity for $L \ne 0$. For $L=0$ there is no zero and so, describes the homogeneous ground state.
Then  one can determine the integrals of Eqs.(\ref{inti2}), (\ref{inti4}) and (\ref{intis}), respectively.
\begin{eqnarray}
I_2&=&\frac{1}{L+1}, \label{I2a}\\
I_4&=&\frac{1}{2L+1}, \, \mbox{and},  \label{I4a}\\
I_{\Upsilon}&=& \frac {4 L} {R^{2}} .\label{Isa}
\end{eqnarray}

%----------------------------------------------
\begin{figure}[h]
\begin{center}
\includegraphics[width=1.0\linewidth]{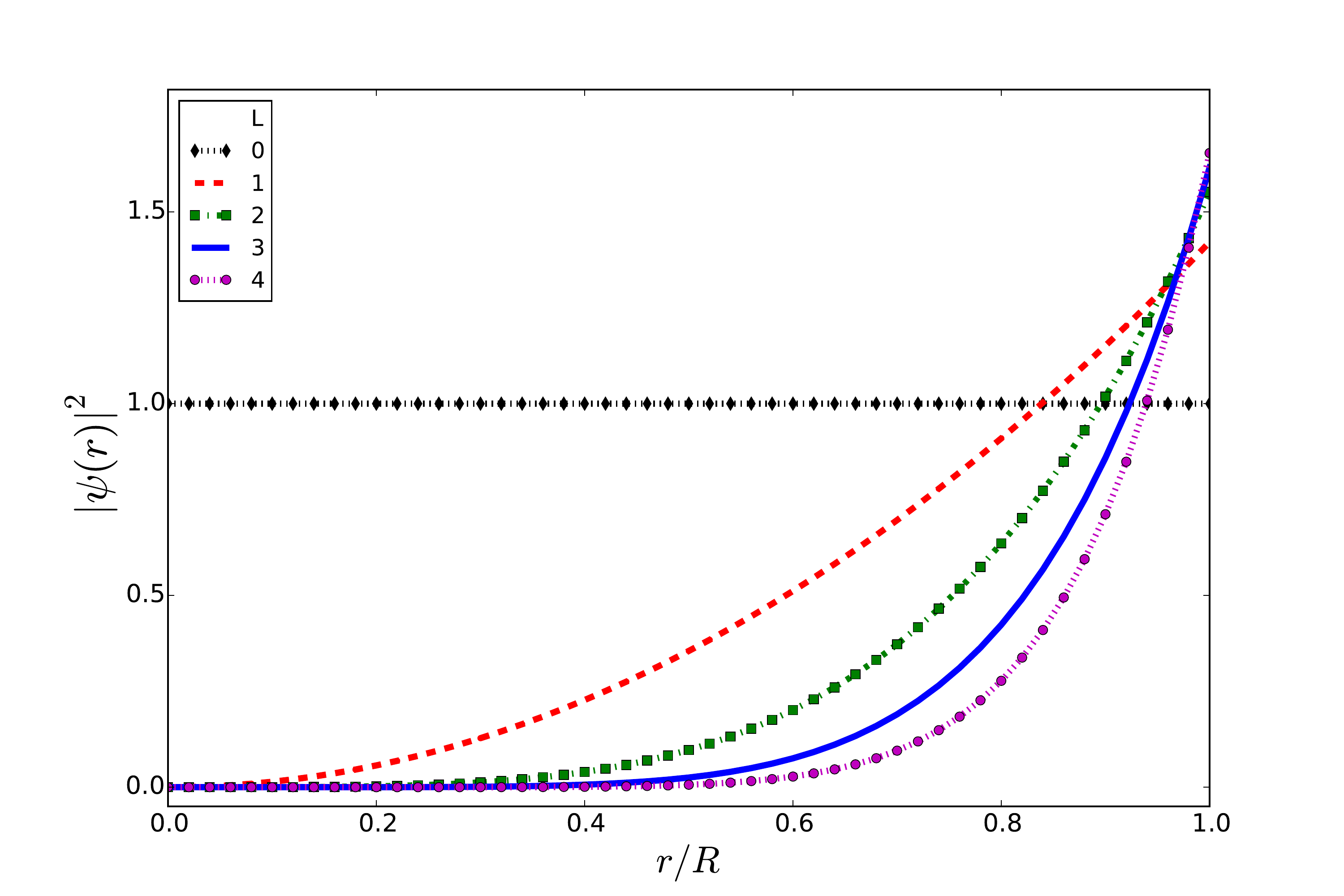}
\caption{The superconducting density as a function of the relative distance to the cylinder's center for vorticity ranging from $L=0$ to $L=4$ in case of a vortex fixed at the center ($\kappa =2.1$).}
\label{psi2vsr-ccv}
\end{center}
\end{figure}
%----------------------------------------------

From Eqs.(\ref{c0PinA}) and (\ref{H*PinA}), one obtains  that,
\begin{eqnarray}
c_{0}^{2}&=&\frac{2\kappa^2 (2L+1)}{2\kappa^2(L+1)+L}, \label{c02VortL}\, \mbox{and}\, , \\	
H'&=&\frac{(2L+1)}{2\kappa^2(L+1)+L}.
\end{eqnarray}
Inserting these expressions into the Gibbs free energy density, either Eq.(\ref{GFEDensc0}) or (\ref{GFEDensH*}), it gives  that,
\begin{eqnarray} \label{gibbs-disk}
\Delta\mathcal{G}&=& -\frac{1}{2}\left [\frac{1}{L+1}-
\frac{4L}{R^2}-\frac{L}{L+1}\frac{(2L+1)}{2\kappa^2(L+1)+L}\right ] \nonumber \\
&\times & \frac{2\kappa^2(2L+1)}{2\kappa^2(L+1)+L}.
\end{eqnarray}
The condition for the existence of the EVS, $\Delta\mathcal{G}\leq 0$, can only be achieved for a limited range of parameters $R$ and $\kappa$. It means that the EVS exists for a given $\kappa$ superconductor in case the radius is larger than a critical value, given by,
\begin{eqnarray}
R_c = 2\sqrt{L (L+1) \cdot \frac{2\kappa^2(L+1)+L}{2\kappa^2(L+1)-2 L^2}. }\label{Rmin}
\end{eqnarray}
In the limit $\kappa \rightarrow \infty$, the critical radius becomes $R_c=2\sqrt{L(L+1)}$, and the Gibbs free energy for $R>R_c$ becomes negative, $\Delta \mathcal{G} \rightarrow [4L(L+1)-R^2](2L+1)/[2 R^2 (L+1)^2]$.
The order parameter and the local magnetic field are given by,
\begin{eqnarray}
\psi(r,\theta)&=& \sqrt{\frac{2\kappa^2 (2L+1)}{2\kappa^2(L+1)+L}}\: \Big(\frac{r}{R}\Big)^L e^{i L \theta}, \, \mbox{and},\\
h_{3}(r)&=&\frac{(2L+1)}{2\kappa^2(L+1)+L}\Big[1-\Big(\frac{r}{R}\Big)^{2L}\Big].
\end{eqnarray}
The paramagnetic magnetization and the field at the center of the vortex, are given by,
\begin{eqnarray}
&& M_3=\frac{L}{L+1}\frac{2\kappa^2(2L+1)}{2\kappa^2(L+1)+L}, \, \mbox{and},\\
&& h_3(v)= \frac{(2L+1)}{2\kappa^2(L+1)+L}.\label{h3v}
\end{eqnarray}
respectively.

%----------------------------------------------
\begin{figure}[h]
\begin{center}
\includegraphics[width=1.0\linewidth]{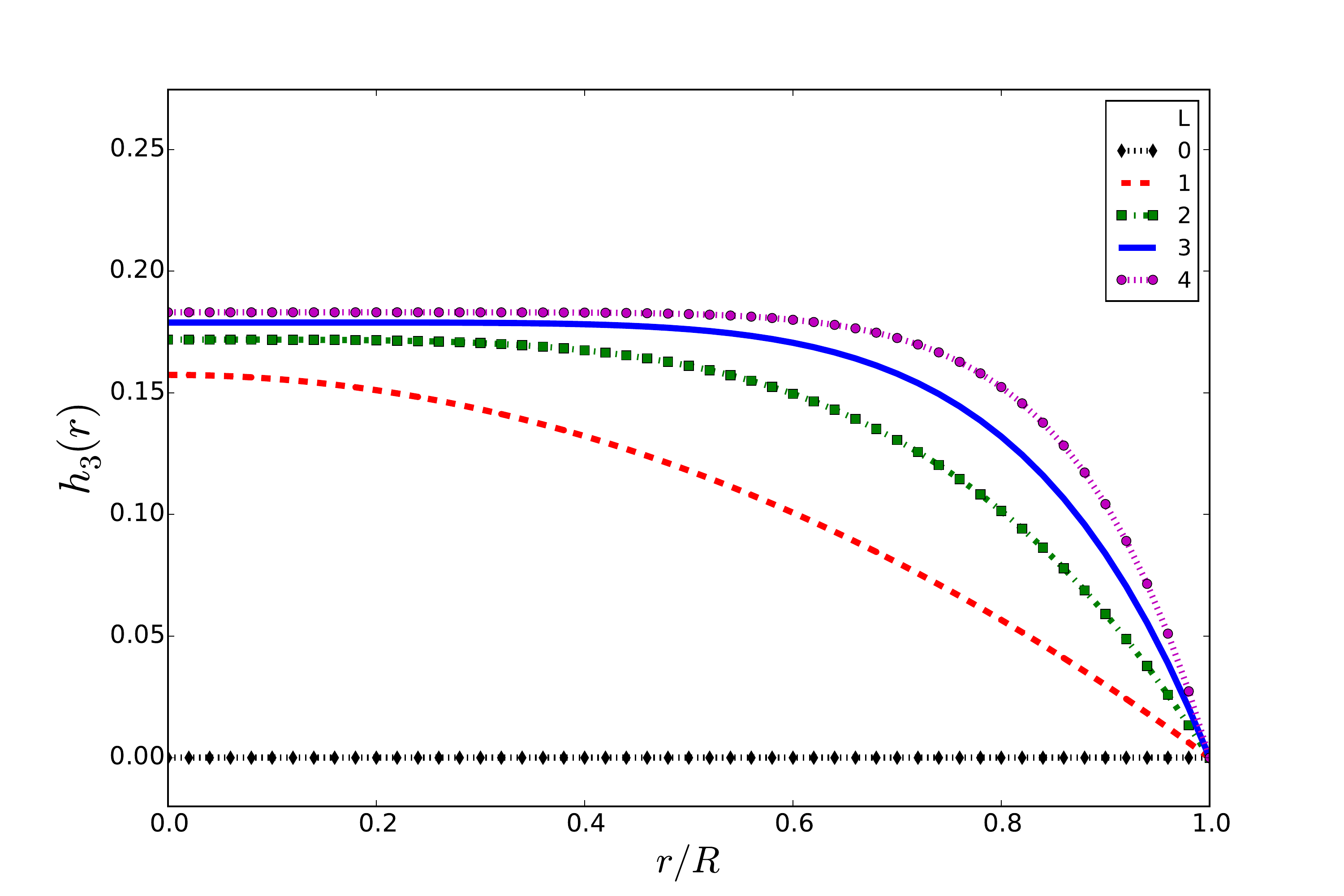}
\caption{The local magnetic field as a function of the relative distance to the cylinder's center for vorticity ranging from $L=0$ to $L=4$ in case of a vortex fixed at the center ($\kappa =2.1$). To recover the Niobium values at the temperature of T = 7.7 K the vertical axis must be multiplied by
$H_{c2}(T=7.7 K)$= 780 Oe}
\label{h3vsr-ccv}
\end{center}
\end{figure}
%----------------------------------------------

\begin{figure}[h]
\begin{center}
\includegraphics[width=1.0\linewidth]{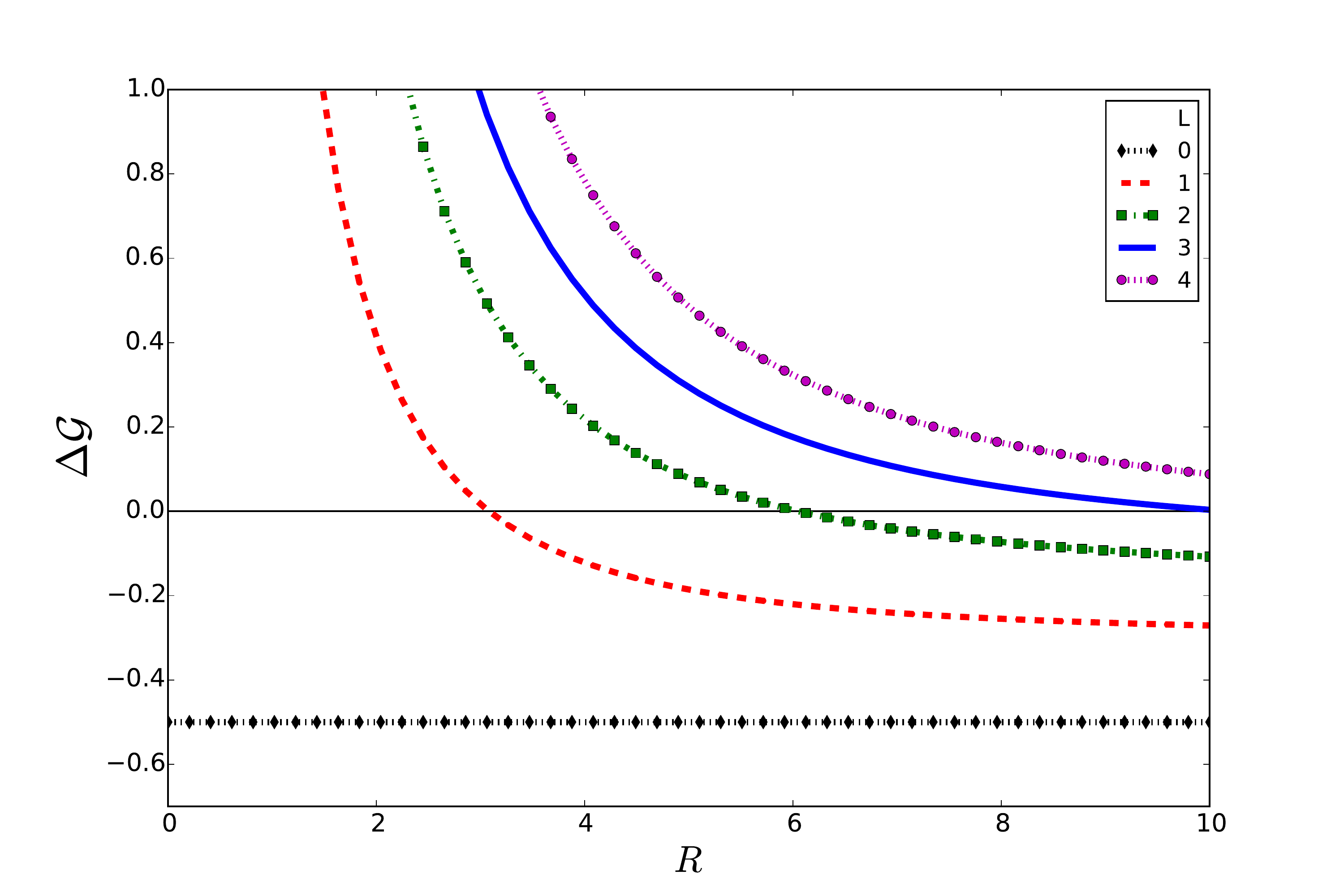}
\caption{The Gibbs free energy difference between the superconducting and the normal states as a function of the cylinder's radius for vorticity ranging from $L=0$ to $L=4$ in case of a vortex fixed at the center ($\kappa =2.1$). The excited vortex state only exists in the negative range of this difference.}
\label{DGibbsvsR-ccv}
\end{center}
\end{figure}

Fig.~\ref{psi2vsr-ccv} shows the superconducting density according to the distance to the center for vorticity  $L=0$ (the homogeneous state) to $L=4$. The density $|\psi|^2$ obtained from Eq. (\ref{psiVortL}) and  $c_{0}^{2}$ given by Eq. (\ref{c02VortL}).
The superconducting density is maximum at the boundary, which implies in a paramagnetic effect, as previously shown. The superconducting density reaches	$c_{0}^{2}$ at the boundary and is a slow growing function of $L$ for $\kappa \ge 1/\sqrt{2}$.
Fig.~\ref{h3vsr-ccv} depicts the local magnetic field versus the distance to the center, for vorticity ranging from $L=0$  to $L=4$.
This plot shows $h_3(r)$ obtained from Eq. (\ref{Hlocal3}), with $c_{0}^{2}$ and $|\psi|^2$ as described in Fig.~\ref{psi2vsr-ccv}.
Notice that the field at the core $h_3(v)$, where $v$ corresponds to $r=0$, follows from Eq.(\ref{h30}). Thus it depends on $L$ through the integrals $I_{2}$ and $I_{4}$, defined by  Eqs. (\ref{I2a}) (\ref{I4a})).
Interestingly the magnetic field at the center of the vortex and the superconducting density at the boundary are directly related to each other, as previously pointed out. With the exception of the homogeneous state, which has no internal magnetic field, for all other states $L\geq1$ the local magnetic field reaches its maximum at the center and vanishes at the boundary, as expected.
For $L\ge 2$ the local field varies slowly from the center to the middle of the cylinder to then abruptly show a strong decay.
The Gibbs free energy density as a function of the cylinder's radius is shown for two cases, $R=1$ and $R=10$, in Fig.~\ref{DGibbsvsR-ccv}.
This plot shows that the homogenous state $L=0$ is the absolute ground state with the minimum energy, $\Delta \mathcal{G} = -0.5$.
Recall that our search is for the EVS, namely for $\Delta \mathcal{G} < 0 $ otherwise $\Delta \mathcal{G} > 0 $ and the superconducting state can somehow decay into the normal state. For instance, the extreme value of this plot, $R=10$, shows that the $L = 0,\, 1,\, 2 $ states are EVS whereas the $L = 3,\,4 $ ones are not. The Gibbs free energy has the following values, $\Delta \mathcal{G} = -0.27,\,-0.11,\,0.003\,$ and $ 0.09 $ for $L = 1,\,2,\,3,\,4$, respectively. The EVS exists for $R > R_c$, $R_{c}=  3.1,\,6.0,\,10.2,\,17.3\,$ for $L$ equal to $1,\,2,\,3,\,4$, respectively.

%--------------------------------------
%------------------------------------------------------

\subsection{Vorticity $L=1$ at any position}\label{offcenter}
The analytical function of $z$ that describes  the order parameter of a single vortex at any position $0 \leq a \leq R $ inside a disk is given by,
\begin{eqnarray}\label{OPPinA}
\psi = c_0 \, \frac{\frac{z}{R}-\frac{a}{R}}{1-\frac{z}{R}\,\frac{\overline{a}}{R}},
\end{eqnarray}
where $c_0$ is the order parameter value at the boundary.
The coordinate $z$ and the position $a$ can be expresses in polar form, $z=r\,e^{i\theta}$ and $a=a_{0}\,e^{i\alpha}$, where we can consider $\alpha=0$ without any loss of generality.
\begin{eqnarray}\label{OPPinAPolar}
\psi = c_0\, R \, \frac{r e^{i\theta}-a_0}{R^2-a_0\,r\,e^{i\theta}}.
\end{eqnarray}
The density of superconducting electrons is given by,
\begin{eqnarray}\label{SCDensPinA}
|\psi|^2 = |c_0|^2\cdot R^2 \, \frac{r^2+a_{0}^{2}-2\,a_0\,r\,\cos(\theta)}{R^4+a_{0}^{2}\,r^2-2\,a_0\,r\,R^2\,\cos(\theta)}.
\end{eqnarray}
The  integrals of Eqs.(\ref{inti2}),(\ref{inti4}) and (\ref{intis}) can be exactly obtained and are given by,
\begin{eqnarray}
I_2& =&2-\frac{R^2}{a_{0}^2}-\left(\frac{R^2}{a_{0}^2}-1\right)^2\ln\left(1-\frac{a_{0}^2}{R^2}\right), \label{I2final}\\
I_4& =&-4\left(\frac{R}{a_{0}}\right)^4+6\left(\frac{R}{a_{0}}\right)^2-1\nonumber\\
&-&4\left(\frac{R}{a_{0}}\right)^2\left[\left(\frac{R}{a_{0}}\right)^2-1\right]^2\ln\left(1-\frac{a_{0}^2}{R^2}\right), \, \mbox{and}, \label{I4final}\\
I_{\Upsilon}& =& \frac{4}{R^2}.\label{Iupsilonfinal}
\end{eqnarray}
Next we take two special limits of these expressions. In the first one  the vortex is just slightly displaced from the center, and the expressions of the previous section must be retrieved. This is the limit $a_0\,\rightarrow\,0$, but to obtained it, the following approximate expansion for the logarithm function must be introduced, $\ln(1-x^2)\approx-x^2-x^{4}/2-x^{6}/3-x^{8}/4-x^{10}/5-x^{12}/6$. Then one obtains the needed approximated expressions for the integrals $I_2$ and $I_4$, respectively
\begin{eqnarray}
I_2 &\approx& \frac{1}{2}+\frac{1}{3}\left(\frac{a_{0}}{R}\right)^2+\frac{1}{12}\left(\frac{a_{0}}{R}\right)^4+\frac{1}{30}\left(\frac{a_{0}}{R}\right)^6\nonumber\\
&+&\frac{1}{60}\left(\frac{a_{0}}{R}\right)^8-\frac{2}{15}\left(\frac{a_{0}}{R}\right)^{10}+\frac{1}{6}\left(\frac{a_{0}}{R}\right)^{12}, \mbox{and},\\
I_4 &\approx& \frac{1}{3}+\frac{1}{3}\left(\frac{a_{0}}{R}\right)^2+\frac{2}{15}\left(\frac{a_{0}}{R}\right)^4+\frac{1}{15}\left(\frac{a_{0}}{R}\right)^6\nonumber\\
&-&\frac{8}{15}\left(\frac{a_{0}}{R}\right)^8+\frac{2}{3}\left(\frac{a_{0}}{R}\right)^{10}.
\end{eqnarray}
From these approximated expressions we can easily verify that
$I_2 = 1/2$ , $I_4=1/3$ and $I_{\Upsilon} = 4/R^2$, which are exactly the Eqs. (\ref{I2a}), (\ref{I4a}) and (\ref{Isa}) for $L=1$.

The other interesting limit is the vortex very near to the boundary of the cylinder. To treat it we change to the coordinate that express the distance of the vortex to the boundary, defined by $y\equiv R - a_{0}$.
Inserting this new parameter into the order parameter we obtain the superconductor's density given by,
\begin{eqnarray}\label{SCDensPinAx}
 |\psi|^2 = c_{0}^{2} \frac{\left(\frac{y}{R}\right)^2+\left(1-\frac{y}{R}\right)^2-
 2\left(1- \frac{y}{R}\right)\frac{r}{R}\cos(\theta)}{1+\left(\frac{r}{R}\right)^2 \left(1-\frac{y}{R}\right)^2 -2\left(1-\frac{y}{R}\right)\frac{r}{R}\cos(\theta)}.\nonumber\\
\end{eqnarray}
The local magnetic field (Eq. (\ref{Hlocal2})) takes into account the density given by Eq.(\ref{SCDensPinAx}).
The expressions for $c_{0}$ and $H'$, given by Eqs.(\ref{c0PinA}) and (\ref{H*PinA}), respectively, are functions of $I_2$ and $I_4$, which in terms of the coordinate $y$ are given by,
\begin{eqnarray}
I_2& =&2-\left(1-\frac{y}{R}\right)^{-2}\nonumber\\
&-&\left[\left(1-\frac{y}{R}\right)^{-2}-1\right]^2\ln\left[1-\left(1-\frac{y}{R}\right)^{2}\right], \, \mbox{and},\label{I2x}\\
I_4& =&-4\left(1-\frac{y}{R}\right)^{-4}+6\left(1-\frac{y}{R}\right)^{-2}-1\nonumber\\
&-&4\left(1-\frac{y}{R}\right)^{-2}\left[\left(1-\frac{y}{R}\right)^{-2}-1\right]^2\ln\left[1-\left(1-\frac{y}{R}\right)^{2}\right].\label{I4x}\nonumber\\
\end{eqnarray}
Similarly the Gibbs free energy density ( Eq.(\ref{GFEDensc0}) or (\ref{GFEDensH*})) is a function of Eqs.(\ref{I2x}) and  (\ref{I4x}), but also of Eq.(\ref{Iupsilonfinal}), which renders it explicitly $R$ dependent.

%------------------------------------------------------------
\begin{figure}[hb]
\begin{center}
\includegraphics[width=1.0\linewidth]{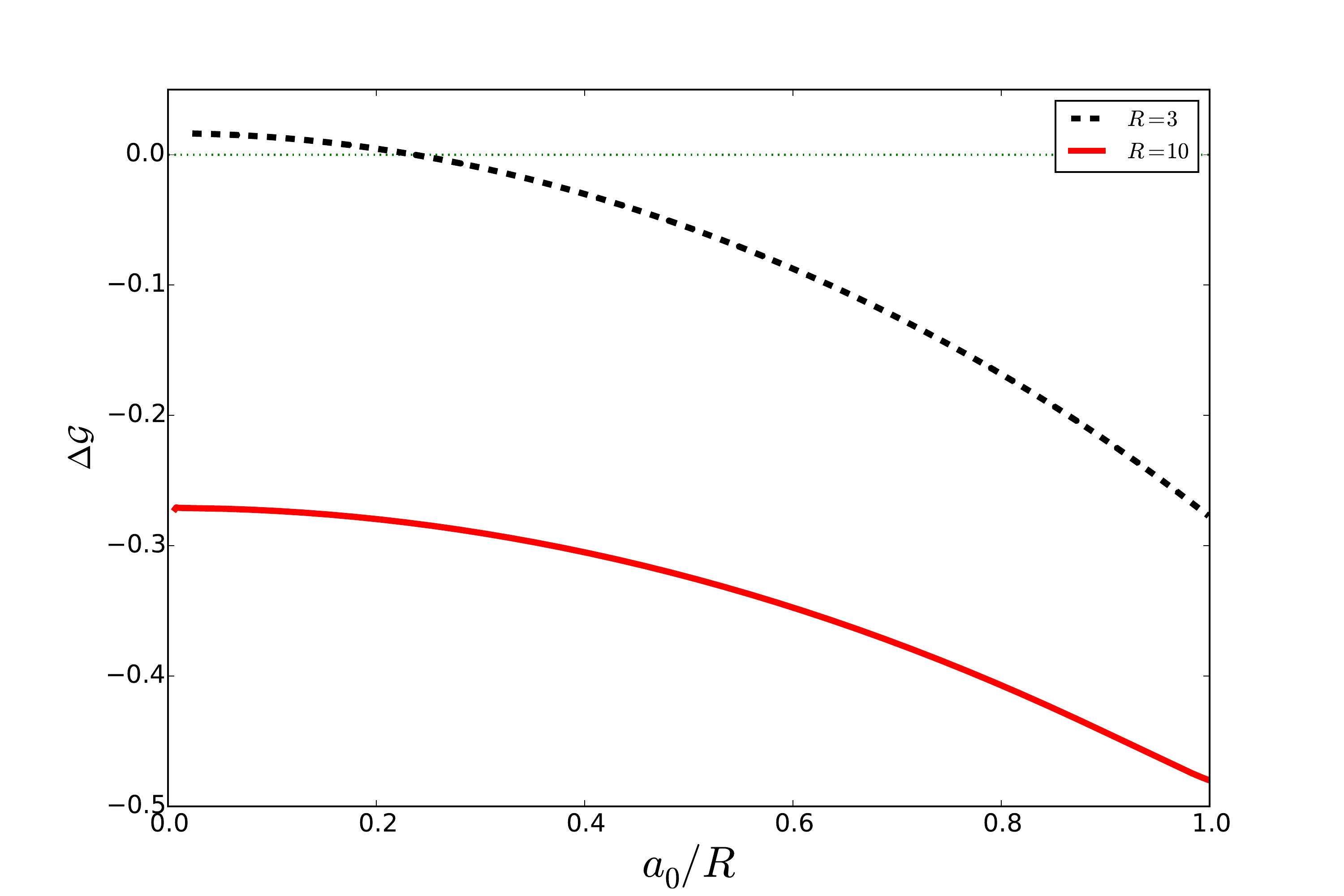}
\caption{The Gibbs free energy difference between the superconducting and the normal states as as a function of the vortex's position inside the cylinder in case of vorticity one. Two radius of cylinders are considered ($\kappa =2.1$). The excited vortex state only exists in the negative range of this difference.}
\label{DGibbsVsaR-cav}
\end{center}
\end{figure}
%--------------------------------------------------

Very near to the boundary of the disk $ R >> y $, which means that the limit  $y\approx 0$ must be taken. Consider only the first order term in the expansion in $y$ to obtain that $I_2 \approx 1 - 2 y / R $ and $I_4 \approx 1 -4 y / R$, and so, $I_2/I_4 \approx 1 + 2 y / R$. Then we  obtain that,
\begin{eqnarray}
&& H' \approx \frac{1}{2 \kappa^2} +  \frac{1}{ \kappa^2}\left(1- \frac{1}{2 \kappa^2}\right)\frac{y}{R}, \, \mbox{and}, \\
&& c_{0}^{2} \approx 1 + 2 \left(1 - \frac{1}{ 2\kappa^2}\right)\frac{y}{R}.
\end{eqnarray}
The paramagnetic magnetization becomes,
\begin{eqnarray}
M_{3} \approx 2 \frac{y}{R}.
\end{eqnarray}
The Gibbs free energy density is given by,
\begin{eqnarray}
\Delta \mathcal{G} \approx -\frac{1}{2}\left(1-\frac{4}{R^2}\right)+\left[ \frac{1}{\kappa^2} +\frac{4}{R^2}\left(1- \frac{1}{2 \kappa^2}\right)\right]\frac{y}{R}.\quad
\end{eqnarray}
For $R \rightarrow \infty$, $c_{0}^{2} = 1$, $ H' = 1/2\kappa^2 $, $M_3=0$ and $\Delta\mathcal{G} = -1/2 $, which means that the homogenous state is recovered.

%--------------------------------------------------
\begin{figure}[hb]
\begin{center}
\includegraphics[width=1.0\linewidth]{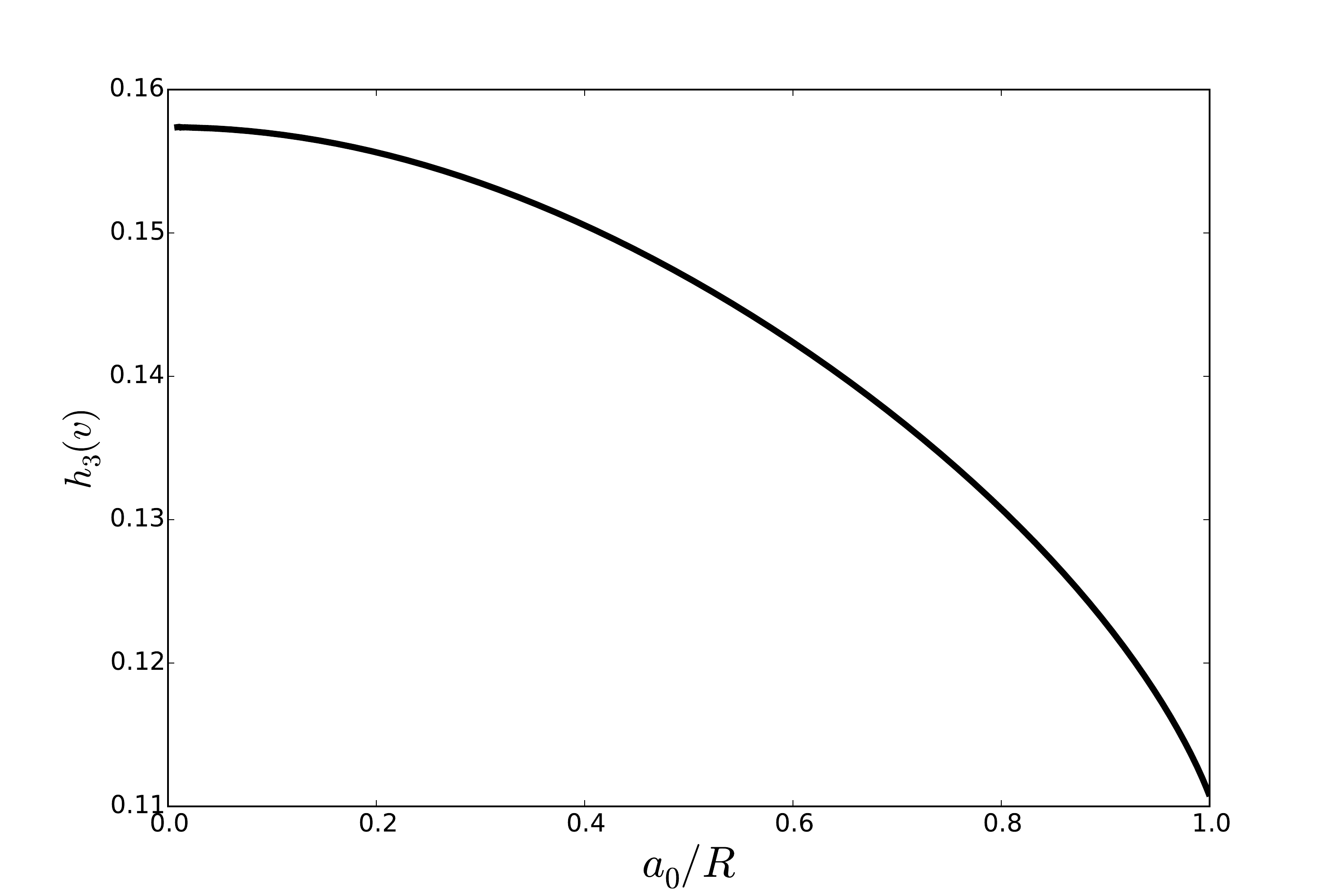}
\caption{The magnetic field at the center of the vortex as a function of the position of the vortex inside the cylinder ($\kappa =2.1$). To recover the value for Niobium at the temperature of T = 7.7 K the vertical axis must be multiplied by $H_{c2}(T=7.7 K)$= 780 Oe.}
\label{h30VsaR-cav}
\end{center}
\end{figure}
%--------------------------------------------------

Fig.~\ref{DGibbsVsaR-cav} shows the Gibbs free energy density with respect to the ratio $a_0/R$.
The plotted curves are obtained from Eq.(\ref{GFEDensc0}) with $c_{0}^{2}$ given by Eq. (\ref{c0PinA}) and the values of the integrals $I_2$, $I_4$ and $I_{\Upsilon}$ given by Eqs. (\ref{I2final}), (\ref{I4final}) and (\ref{Iupsilonfinal}), respectively.
The two cases shown, $R=3$ and $R=10$ demonstrate that the existence of EVS depends on the position of the vortex.
In both cases the Gibbs free energy decreases monotonically from the center to the boundary of the cylinder.
For $R=10$ the vortex in any position is in a EVS, since the Gibbs free energy is always negative, but this is not so for the $R=3$ cylinder. There the Gibbs free energy is positive  for $ a_{0} < 0.24 R $ such that only for $ a_{0} > 0.24 R $ there is an EVS. The Gibbs free energy is null for $a_{0} = 0.24 R$ and decreases to reach the value $\Delta \mathcal{G} = -0.27$ at the boundary.

%---------------------------------------------------------
\begin{figure}[hb]
\begin{center}
\includegraphics[width=1.0\linewidth]{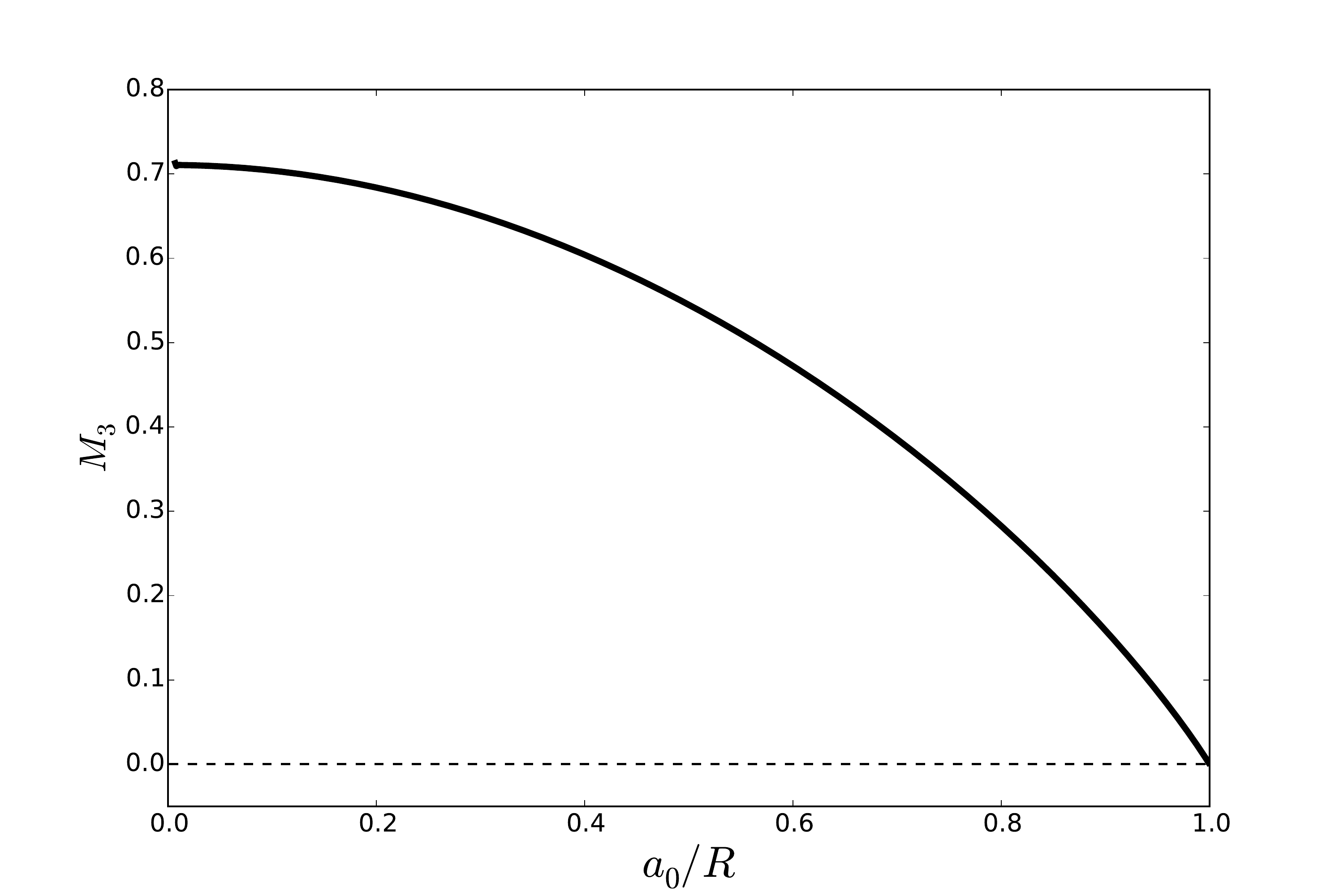}
\caption{The magnetization as a function of the position of the vortex inside the cylinder ($\kappa =2.1$). To recover the value for Niobium for the temperature of T = 7.7 K the vertical axis must be multiplied by
$H_{c2}(T=7.7 K)/8\pi \kappa^2$= 7.0 Oe.}
\label{M3VsaR-cav}
\end{center}
\end{figure}
%---------------------------------------------------------

\begin{figure}[hb]
\begin{center}
\includegraphics[width=1.0\linewidth]{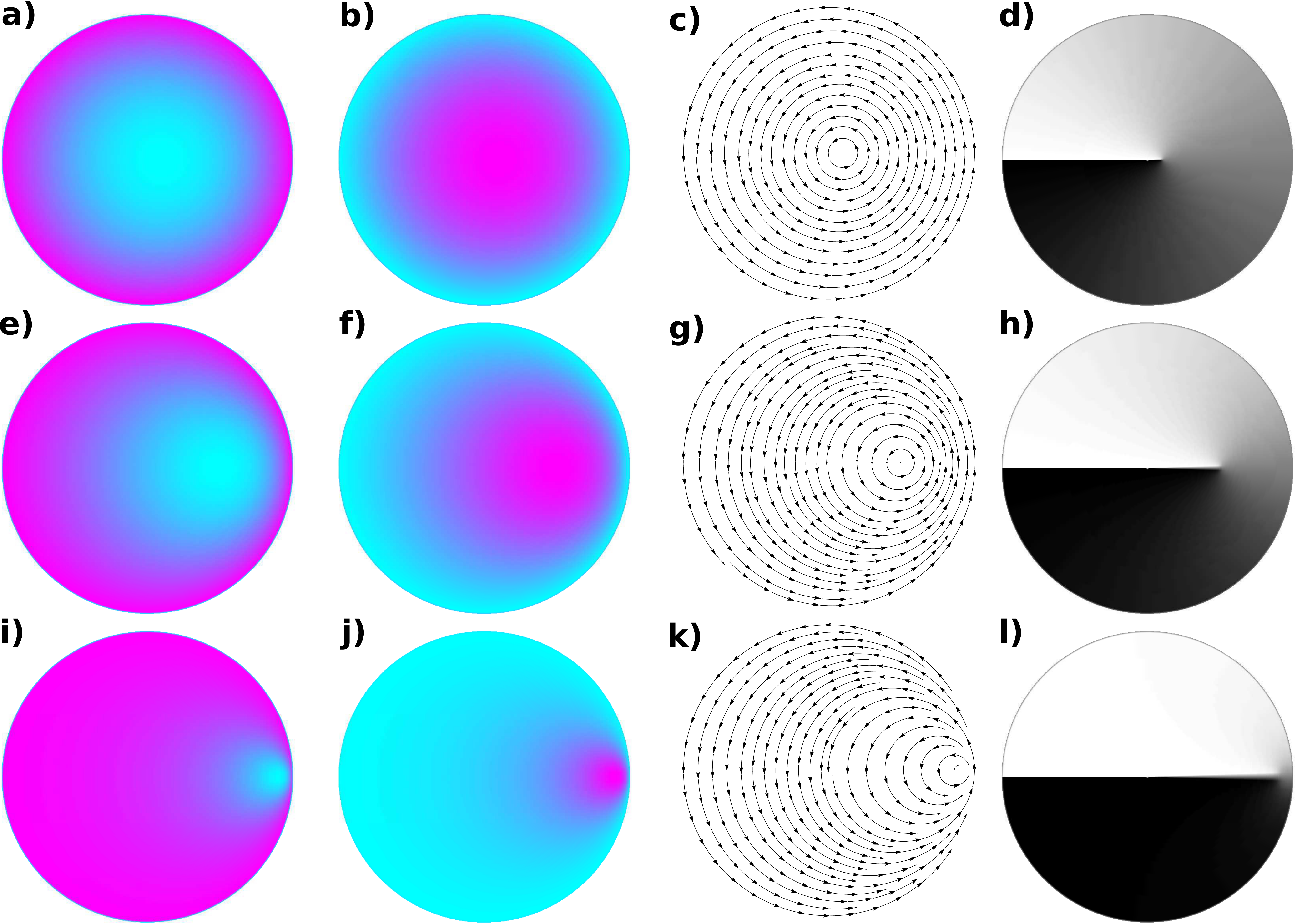}
\caption{The superconducting density, the local magnetic field, the electric current density and the order parameter phase are depicted at three different positions of the vortex inside the cylinder, $a_{0}/ R=0.1$ ((a) to (d)), $0.5$  ((e) to (h)), and $0.9$ ((i) to (l)) of the vortex inside the cylinder.
Figures (a), (e) and (i) display the density in (color on line) scheme ranging from low density (cyan) to high density (magenta).
The (color on line) scheme is distinct for each of these three figures, as the density varies from zero at the center of the vortex to $c_0^2$ at the boundary, whose value varies according to the position of the vortex inside the cylinder (see Eqs.(\ref{SCDensPinA}), (\ref{c0PinA}), (\ref{I2x}) and (\ref{I4x})). The local magnetic field is shown in figures (b), (f) and (j), and the color scheme is
distinct for each of these three figures. It
ranges from the maximum at the center of the vortex (magenta), according to Eqs.(\ref{Hlocal2}) and (\ref{h3v}), to zero at the boundary (cyan). Figures (c), (g) and (k) show the vortex current around the vortex obtained from Eq.(\ref{curr-foe}). Figures (d), (h) and (l) show the phase of the order parameter. The discontinuity line splitting the white to the black region shows  the existence of a single vortex inside the cylinder. The color range from white to black describes $0$ to $2\pi$ phase. }
\label{figcolor}
\end{center}
\end{figure}

Fig.~\ref{h30VsaR-cav} illustrates how the local magnetic field at the vortex's core varies according to the vortex position.
This figure shows that this field  $h_{3} (v) $  depends on the ratio $ a_{0}/R$ according to Eq. (\ref{h30}). Thus it only depends on the ratio $I_{2}/I_{4} $, defined by Eqs. (\ref{I2final}) and (\ref{I4final}), respectively.
The local field is maximum at the vortex's center. From its side $h_3(v)$ reaches a maximum when the vortex is at the center and slowly decreases when the vortex is positioned at the boundary.
The Fig.~\ref{h30VsaR-cav} is in agreement with Fig.~\ref{h3vsr-ccv} since
$h_{3} (0)$ for $a_{0} =0 $ corresponds to $h_3 (r = 0 ) $ for $ L= 1$.
Fig.~\ref{M3VsaR-cav} shows the paramagnetic magnetization as a function of the ratio $a_{0}/ R $ as obtained from Eq. (\ref{M3}).
Interestingly the magnetization  depends on the position of the vortex, according to the integrals $I_{2}$ and $I_{4}$, given by Eqs.(\ref{I2final}) and (\ref{I4final}), respectively.
The magnetization is stronger for the vortex near to the cylinder's center and weaker near to the cylinder's border.
For a vortex at the boundary, $a_{0}=R$, the magnetization vanishes because $ I_{2} = 1 $. This is consistent with the description of the exit of the vortex, although $h_{3}(v)$ is not zero as can be seen in Fig.~\ref{h30VsaR-cav}.
The vortex next to the boundary means that the superconducting density is almost homogenous on the entire region of the cylinder and the integral $I_{2}$ is almost equal to the unity what makes the magnetization goes to zero as can be seen from Eq.(\ref{M3}). Finally the three rows of Fig.~\ref{figcolor} depict the vortex at three different positions, $a_{0}/R=0.1$, $0.5$ and $0.9$, respectively. The columns correspond to the density, local magnetic field, current and phase, respectively. The density is obtained from Eqs.(\ref{SCDensPinA}) and (\ref{c0PinA}), with the integrals of Eqs.(\ref{I2x}) and (\ref{I4x}).
Therefore the density ranges from zero at the center of the vortex to the maximum value $c_0^2$ at the boundary. The density is shown in Figs.~\ref{figcolor}(a), (e) and (i) in color (on line) scheme ranging from low density (cyan) to high density (magenta). The local magnetic field is obtained from Eq.(\ref{Hlocal2}) and ranges from the value given by Eq.(\ref{h3v}) at the center of the vortex to zero at the boundary. Thus the (color on line) scheme of Figs.~\ref{figcolor}(b), (f) and (j) ranges from the maximum at the center of the vortex (magenta) to zero at the boundary(cyan). The vortex current is obtained from Eq.(\ref{curr-foe}) and depicted in Figs.~\ref{figcolor}(c), (g) and (k). Notice that there is no current flowing out of the cylinder as the vortex moves towards the boundary, which is a general property guaranteed by the present formalism.
Finally Figs.~\ref{figcolor}(d), (h) and (l) show the phase of the order parameter, defined as $\tan^{-1}\left (\psi_I/\psi_R \right )$. The discontinuity, represented as a straight line that abruptly separates white to black ,confirms the presence of a vortex in the cylinder.
%------------------------------------------------------------
%------------------------------------------------------------

\section{Conclusion} \label{conclusion}
In this paper we have included the treatment of boundaries in the first order equations method used by A. Abrikosov to predict the vortex lattice. We find a direct connection between this method and the theory of conformal mapping.
Using this method we have considered vortex states left inside the superconductor after the external applied field is switched off.
This vortex state is unstable which means that vortices must leave the superconductor since their presence is thermodynamically forbidden. However vortices are topological stable as long as the superconducting state exists.
We define here the excited vortex state, whose free energy is lower than that of the normal state.
Thus the excited vortex state cannot collapse into the normal state.
We calculate the paramagnetic magnetization of the excited vortex state for some particular cases in a thin long wires of Niobium.

\textbf{Acknowledgments}:
We acknowledge J. C. Lopez Ortiz for helpful discussions. M.M.D acknowledges CNPq (Brazil) support from funding (23079.014992/2015-39). R. R. G. acknowledges CNPq (Brazil) support from funding (207007/2014-4). A. R. de C. R.  acknowledges FACEPE (Brazil) support from funding (APQ-0198-1.05/14).

\appendix
\section{The kinetic energy decomposition} \label{appendix_a}
Consider the term $|D_{+}\psi|^{2}$, which can be casted as,
\begin{eqnarray}
|D_{+}\psi|^{2}&=&[(D_{1}\psi)^*-i(D_{2}\psi)^*][(D_{1}\psi)+i(D_{2}\psi)]\quad \nonumber \\
&=&(D_1\psi)^*(D_1\psi)+(D_2\psi)^*(D_2\psi)\nonumber\\
&+&i[(D_1\psi)^*(D_2\psi)-(D_2\psi)^*(D_1\psi)].
\end{eqnarray}
Expanding only the derivative $D_1$,
\begin{eqnarray}
(D_1\psi)^*(D_2\psi)&=&\Big( -\frac{\hbar}{i}\partial_1\psi^* \nonumber
\frac{q}{c}A_1\psi^*\Big)(D_2\psi)\nonumber\\
&=&\partial_1\Big[-\frac{\hbar}{i}\psi^*(D_2\psi)\Big]\nonumber\\
&-&(\frac{\hbar}{i}\psi^*)\partial_1 D_2 \psi - \frac{q}{c}A_1\psi^*D_2\psi.
\end{eqnarray}
Rearranging the terms, one obtains that,
\begin{eqnarray}
(D_1\psi)^*(D_2\psi)=-\frac{\hbar}{i}\partial_1(\psi^*D_2\psi)+\psi^*D_1D_2\psi.
\end{eqnarray}
In the same way, we obtain that,
\begin{eqnarray}
(D_2\psi)^*(D_1\psi)=
-\frac{\hbar}{i}\partial_2(\psi^*D_1\psi)+\psi^*D_2 D_1\psi.
\end{eqnarray}
The complex conjugate of these relations give us that,
\begin{eqnarray}
&&(D_2\psi)^*(D_1\psi)=
\frac{\hbar}{i}\partial_1[(D_2\psi)^*\psi]+(D_1D_2\psi)^*\psi, \, \\
&&(D_1\psi)^*(D_2\psi)= \frac{\hbar}{i}\partial_2[(D_1\psi)^*\psi+(D_2
D_1\psi)^*\psi.
\end{eqnarray}
At this point, we have two formulations for the same identity, where one relation is the complex conjugate of the second one. The expression
\begin{eqnarray}
&&|D_+\psi|^2=|\vec{D}\psi|^2\nonumber\\
&+&i\Big[-\frac{\hbar}{i}\partial_1(\psi^*D_2\psi)+\psi^*D_1D_2\psi\Big]\nonumber\\
&-&\Big[-\frac{\hbar}{i}\partial_2(\psi^*D_1\psi)+\psi^*D_2D_1\psi\Big]\nonumber,
\end{eqnarray}
becomes,
\begin{eqnarray}
|D_+\psi|^2&=&|\vec{D}\psi|^2+i\psi^*[D_1,D_2]\psi\nonumber\\
&-&\hbar\partial_1(\psi^*D_2\psi)+\hbar\partial_2(\psi^*D_1\psi).
\end{eqnarray}
The commutator,
\begin{eqnarray}
[D_1,D_2]&=&[\frac{\hbar}{i}\partial_1-\frac{q}{c}A_1,\frac{\hbar}{i}\partial_2-\frac{q}{c}A_2]\nonumber\\
&=&-\frac{\hbar}{i}\frac{q}{c}(\partial_1A_2-\partial_2A_1),
\end{eqnarray}
becomes,
\begin{eqnarray}
[D_1,D_2]=-\frac{\hbar q}{i c} h_3.
\end{eqnarray}
Then one obtains that,
\begin{eqnarray}
|D_+\psi|^2&=&|\vec{D}\psi|^2-i([D_1,D_2]\psi^*)\psi\nonumber\\
&-&\hbar\partial_1[(D_2\psi)^*\psi]+\hbar\partial_2{(D_1\psi)^*\psi]},
\end{eqnarray}
which gives  expressions for $|\vec{D}\psi|^2 $,
\begin{eqnarray}
|\vec{D}\psi|^2&=&|D_+\psi|^2+\frac{\hbar q}{c} h_3|\psi|^2\nonumber\\
&+&\hbar[\partial_1(\psi^*D_2\psi)-\partial_2(\psi^*D_1\psi)],
\end{eqnarray}
and
\begin{eqnarray}\label{apwl}
|\vec{D}\psi|^2&=&|D_+\psi|^2+\frac{\hbar q}{c} h_3|\psi|^2\nonumber\\
&+&\hbar\{\partial_1[(D_2\psi)^*\psi]-\partial_2[(D_1\psi)^*\psi]\}.
\end{eqnarray}
We sum the two expressions and divide by 2 to obtain that,
\begin{eqnarray}
|\vec{D}\psi|^2&=&|D_+\psi|^2+\frac{\hbar q}{c} h_3 |\psi|^2\nonumber\\
&+&\hbar\{\partial_1\frac{\psi^*D_2\psi+(D_2\psi)^*\psi}{2}-\partial_2\frac{\psi^*D_1\psi+(D_1\psi)^*\psi}{2}\}.\nonumber\\
\end{eqnarray}
Introducing the definition of the current we obtain the desired dual formulation of the kinetic energy given by Eq.(\ref{math-iden})

\section{Calculation of the integrals}\label{appendix_b}
In this appendix we calculate the following integrals.
\begin{eqnarray}
&&I_2=\frac{R^2}{\Sigma}\int_{0}^{R}dr\,r\,f_2(r)\label{DefI2},\\
&&f_2(r)=\int_{0}^{2\pi}d\theta\frac{r^2+a_{0}^{2}-2\,a_0\,r\,\cos(\theta)}{R^4+a_{0}^{2}\,r^2-2\,a_0\,r\,R^2\,\cos(\theta)}\label{Deff2r},\\
&&I_4=\frac{R^4}{\Sigma}\int_{0}^{R}dr\,r\,f_4(r)\label{DefI4},\\
&&f_4(r)=\int_{0}^{2\pi}d\theta\frac{\left(r^2+a_{0}^{2}-
2\,a_0\,r\,\cos(\theta)\right)^2}{\left(R^4+a_{0}^{2}\,r^2-2\,a_0\,r\,R^2\,\cos(\theta)\right)^2}\label{Deff4r},\\
&&I_{\Upsilon}=\frac{1}{\Sigma}\int d^2x\,\vec{\partial}^2 f_{\Upsilon}(r,\theta)= \frac{1}{\Sigma}\oint_{\partial\Sigma}dl \,\hat{n}\cdot\vec{\partial}f_{\Upsilon}(r,\theta)\label{DefIupsilon},\\
&&f_{\Upsilon}(r,\theta)=R^2\frac{r^2+a_{0}^{2}-2\,a_0\,r\,\cos(\theta)}{R^4+a_{0}^{2}\,r^2-2\,a_0\,r\,R^2\,\cos(\theta)}\label{Deffupsilon}.
\end{eqnarray}
The integral $I_2$ is associated with the density of Cooper pairs.
As shown in Eq.(\ref{DefI2}) the integral of $|\psi|^2$ is obtained in two parts: first we integrate the angular part, that results in a function of the radius $f_2(r)$ and after we integrate the function $f_2 (r)$ in the radial coordinate.
The angular integral has a form with a well-known result~\cite{gradshteyn14}.
\begin{eqnarray}\label{Generalformf2r}
 &&\int dx \frac{A+B\;cos(x)}{a+b\;cos(x)}=\frac{B}{b}x+\frac{A\;b-a\;B}{b}\int dx \frac{1}{a+bcos(x)},\nonumber\\
 &&\\
 &&\int dx \frac{1}{a+b cos(x)}=\frac{2}{\sqrt{a^2-b^2}} \tan^{-1} \left ( \frac{\sqrt{a^2-b^2}\tan(x/2)}{a+b}\right)\nonumber\\
 &&\mbox{, for $a^2\;>\;b^2$.}
\end{eqnarray}
To calculate this integral we consider separately the integration in the four quadrants,as shown below.
\begin{eqnarray}\label{f2rquadrant}
 &&\int_{0}^{2 \pi} dx \frac{A+B cos(x)}{a+b cos(x)}=\int_{0}^{\pi/2} dx \frac{A+B cos(x)}{a+b cos(x)}\nonumber\\
 &+&\int_{\pi/2}^{\pi} dx \frac{A+B cos(x)}{a+b cos(x)}+\int_{\pi}^{3 \pi/2} dx \frac{A+B cos(x)}{a+b cos(x)}\nonumber \\
 &+&\int_{3 \pi /2}^{2 \pi} dx \frac{A+B cos(x)}{a+b cos(x)}.
\end{eqnarray}
We make the following change of variables.

First quadrant: $ y = x $
\begin{eqnarray}
 \int_{0}^{\pi/2} dx \frac{A+B cos(x)}{a+b cos(x)} \rightarrow \int_{0}^{\pi/2} dy \frac{A+B cos(y)}{a+b cos(y)}.\nonumber\\
\end{eqnarray}

Second quadrant: $ y = x-\pi $
\begin{eqnarray}
 \int_{\pi/2}^{\pi} dx \frac{A+B cos(x)}{a+b cos(x)}\rightarrow \int_{-\pi/2}^{0} dy \frac{A-B cos(y)}{a-b cos(y)}.\nonumber\\
\end{eqnarray}

Third quadrant: $ y = x-\pi $
\begin{eqnarray}
 \int_{\pi}^{3\pi/2} dx \frac{A+B cos(x)}{a+b cos(x)} \rightarrow \int_{0}^{\pi/2} dy \frac{A-B cos(y)}{a-b cos(y)}.\nonumber\\
\end{eqnarray}

Fourth quadrant: $ y = x-2\pi $
\begin{eqnarray}
 \int_{3\pi/2}^{2\pi} dx \frac{A+B cos(x)}{a+b cos(x)} \rightarrow \int_{-\pi/2}^{0} dy \frac{A+B cos(y)}{a+b cos(y)}.\nonumber\\
\end{eqnarray}

Inserting the variable changes in the original integral, we obtain a new interval of integration
\begin{eqnarray}
 &&\int_{0}^{2 \pi} dx \frac{A+B cos(x)}{a+b cos(x)}=\int_{-\pi/2}^{\pi/2} dy \frac{A+B cos(y)}{a+b cos(y)}\nonumber\\
 &+&\int_{-\pi/2}^{\pi/2} dy \frac{A-B cos(y)}{a-b cos(y)}.
\end{eqnarray}
We calculate the integrals in the interval $ -\pi/2 < x < + \pi/2 $
\begin{eqnarray}
 &&\int_{-\pi/2}^{+\pi/2} dx \frac{A+B cos(x)}{a+b cos(x)}=\frac{B}{b} \pi \nonumber\\
 &+&\frac{4(Ab-aB)}{b\sqrt{a^2-b^2}} \tan^{-1} \left ( \frac{\sqrt{a^2-b^2}}{a+b}\tan(\frac{\pi}{4})\right),
\end{eqnarray}

\begin{eqnarray}
 &&\int_{-\pi/2}^{+\pi/2} dx \frac{A-B cos(x)}{a-b cos(x)}=\frac{B}{b} \pi \nonumber\\
 &+&\frac{4(A\;b-a\;B)}{b\sqrt{a^2-b^2}} \tan^{-1} \left ( \frac{\sqrt{a^2-b^2}}{a-b}\tan(\frac{\pi}{4})\right).
\end{eqnarray}

We obtain the expressions for the integral in the interval $0 < x < 2\pi$
\begin{eqnarray}
 &&\int_{0}^{2\pi} dx \frac{A+B cos(x)}{a+b cos(x)}=2\pi\frac{B}{b} + \frac{4(A\;b-a\;B)}{b\sqrt{a^2-b^2}}\nonumber\\
 &&\times\left [ \tan^{-1} \left ( \frac{\sqrt{a^2-b^2}}{a+b}\right)+ \tan^{-1} \left ( \frac{\sqrt{a^2-b^2}}{a-b}\right)\right],
\end{eqnarray}
and verify that,
\begin{eqnarray}\nonumber
\frac{\sqrt{a^2-b^2}}{a+b}=\frac{\sqrt{(a+b)(a-b)}}{(a+b)^2}=\left(\frac{a-b}{a+b}\right)^{1/2 },\nonumber\\
\frac{\sqrt{a^2-b^2}}{a-b}=\frac{\sqrt{(a+b)(a-b)}}{(a-b)^2}=\left(\frac{a+b}{a-b}\right)^{1/2 }.\nonumber\\
\end{eqnarray}
Using the identity $ tan^{-1}(x)+tan^{-1}(1/x)=\pi/2 $ if  $  x>0 $, we simplify the expression to obtain that, \begin{eqnarray}\label{GeneralResultf2r}
 \int_{0}^{2\pi} dx \frac{A+B cos(x)}{a+b cos(x)}=2\pi\frac{B}{b}+2\pi\frac{(A\;b-a\;B)}{b\sqrt{a^2-b^2}}.\nonumber\\
\end{eqnarray}
At this point we have a general result for the integral (\ref{Generalformf2r}).
The next step is to apply this result for integrand of $f_{2}(r)$.
We consider the following identifications $A=a_{0}^{2} + r^{2}$, $B=-2a_{0}r$, $a=R^{4}+a_{0}^{2}r^{2}$ and $b=-2 a_{0} r R^{2}$.
Inserting these expressions in the Eq. (\ref{GeneralResultf2r}) we obtain
\begin{eqnarray}\label{f2r}
 f_2 (r) = \frac{2\pi}{R^2}-\frac{2\pi}{R^2}\frac{(R^2-a_{0}^{2})(R^2-r^2)}{R^4 - a_{0}^{2}r^2}.
\end{eqnarray}
With the  result obtained in Eq.(\ref{f2r}), the integral $I_2$ becomes,
\begin{eqnarray}\label{I2intdr}
 I_2 &=& \frac{R^2}{\Sigma}\int_{0}^{R}r f_2 (r) dr\nonumber\\
 &=& \frac{2\pi}{\Sigma} \int_{0}^{R} r dr - \frac{2\pi}{\Sigma}(R^2 - a_{0}^{2})R^2 \int_{0}^{R} \frac{r dr}{R^4 - a_{0}^{2}r^{2}}\nonumber\\
 &+& \frac{2\pi}{\Sigma} (R^2 - a_{0}^{2}) \int_{0}^{R} \frac{r^3 dr}{R^4 -a_{0}^{2}r^{2}}.
\end{eqnarray}
Next we calculate each one of the three integrals in Eq.(\ref{I2intdr})
\begin{eqnarray}
 &&\int_{0}^{R} r dr = \frac{R^2}{2},\nonumber\\
 &&\int_{0}^{R} \frac{r dr}{R^4 - a_{0}^{2}r^{2}}=\frac{1}{2 a_{0}^{2}}\ln\left(\frac{R^4}{R^4 - a_{0}^{2}R^{2}}\right),\nonumber\\
 &&\int_{0}^{R} \frac{r^3 dr}{R^4 -a_{0}^{2}r^{2}}=\frac{R^4}{2 a_{0}^{4}}\ln\left(\frac{R^4}{R^4 - a_{0}^{2}R^{2}}\right)-\frac{R^2}{R^2 - a_{0}^{2}}.\nonumber
\end{eqnarray}
Inserting these results in Eq.(\ref{I2intdr}) and considering that $\Sigma = \pi R^2$ we have
\begin{eqnarray}
 I_2 = 2- \frac{R^2}{a_{0}^{2}}+\left( \frac{R^2}{a_{0}^{2}}-1 \right)^2\ln\left( 1 - \frac{a_{0}^{2}}{R^{2}} \right).
\end{eqnarray}

Next we calculate the surface integral defined in Eq.(\ref{DefIupsilon}).
The normal vector is the radial one, $\hat{n} = \hat{r}$, such that, $f_{\Upsilon}(r,\theta)$:
$\hat{r}\cdot\vec{\partial}f_{\Upsilon}= \partial_{r} f_{\Upsilon}$.
To calculate the integral with the derivative,
\begin{eqnarray}
 &&\frac{\partial f_{\Upsilon}(r,\theta)}{\partial r}\nonumber\\
 &=& 2 R^2 (R^2 - a_{0}^{2})\frac{(R^2 + a_{0}^{2})r - a_{0}(R^2 + r^{2})\cos(\theta)}{(R^4 +  a_{0}^{2}r^2 - 2 a_{0} r R^2 \cos(\theta))^2 },\nonumber\\
\end{eqnarray}
we use the formula~\cite{gradshteyn14}
\begin{eqnarray}
 \int dx \frac{A+B \cos(x)}{(a+b \cos(x))^2}&=&\frac{1}{a^2 - b^2}\left[\frac{(a\;B-A\;b)\sin(x)}{a+b\cos(x)}\right.\nonumber\\
 &+&\left.\int dx\frac{A\;a-b\;B}{a+b\cos(x)}\right].
\end{eqnarray}
To obtain the integral $I_{\Upsilon}$, we divide the interval $0 < x < 2\pi$ into four quadrants and use the following changes of variable for each of them.

First quadrant: $ y = x $
\begin{eqnarray}
 \int_{0}^{\pi/2} dx \frac{A+B \cos(x)}{(a+b \cos(x))^2} \rightarrow \int_{0}^{\pi/2} dy \frac{A+B \cos(y)}{(a+b \cos(y))^2}.\nonumber\\
\end{eqnarray}

Second quadrant: $ y = x-\pi $
\begin{eqnarray}
 \int_{\pi/2}^{\pi} dx \frac{A+B \cos(x)}{(a+b \cos(x))^2}\rightarrow \int_{-\pi/2}^{0} dy \frac{A-B \cos(y)}{(a-b \cos(y))^2}.\nonumber\\
\end{eqnarray}

Third quadrant: $ y = x-\pi $
\begin{eqnarray}
 \int_{\pi}^{3\pi/2} dx \frac{A+B \cos(x)}{(a+b \cos(x))^2} \rightarrow \int_{0}^{\pi/2} dy \frac{A-B \cos(y)}{(a-b \cos(y))^2}.\nonumber\\
\end{eqnarray}

Fourth quadrant: $ y = x-2\pi $
\begin{eqnarray}
 \int_{3\pi/2}^{2\pi} dx \frac{A+B \cos(x)}{(a+b \cos(x))^2} \rightarrow \int_{-\pi/2}^{0} dy \frac{A+B \cos(y)}{(a+b \cos(y))^2}.\nonumber\\
\end{eqnarray}
We collect these results together and we obtain a new integration interval $-\pi/2 < x < \pi/2$, and the original integral is rewritten in two parts, with
positive and negative signs in the integrand, respectively.
\begin{eqnarray}
 &&\int_{0}^{2 \pi} dx \frac{A+B \cos(x)}{(a+b \cos(x))^2}=\int_{-\pi/2}^{\pi/2} dy \frac{A+B \cos(y)}{(a+b \cos(y))^2}\nonumber\\
 &+&\int_{-\pi/2}^{\pi/2} dy \frac{A-B \cos(y)}{(a-b \cos(y))^2}.
\end{eqnarray}
For the integral with the positive sign we find that,
\begin{eqnarray}
 &&\int_{-\pi/2}^{\pi/2} dy \frac{A+B \cos(y)}{(a+b \cos(y))^2}=\frac{2(a\;B-A\;b)}{a(a^2-b^2)}\nonumber\\
 &+&\frac{4(A\;a-b\;B)}{(a^2-b^2)^{3/2}}\tan^{-1}\left( \frac{\sqrt{a^2-b^2}}{a+b} \right),
\end{eqnarray}
and for the negative sign,
\begin{eqnarray}
 &&\int_{-\pi/2}^{\pi/2} dy \frac{A-B \cos(y)}{(a-b \cos(y))^2}=-\frac{2(a\;B-A\;b)}{a(a^2-b^2)}\nonumber\\
 &+&\frac{4(A\;a-b\;B)}{(a^2-b^2)^{3/2}}\tan^{-1}\left( \frac{\sqrt{a^2-b^2}}{a-b} \right).
\end{eqnarray}
Summing the two integrals and using the same trigonometric property for $tan^{-1}(x)$ shown before, we obtain that,
\begin{eqnarray}\label{IntCos0_2pi}
 &&\int_{0}^{2 \pi} dx \frac{A+B \cos(x)}{(a+b \cos(x))^2}=2\pi \frac{A\;a-b\;B}{(a^2-b^2)^{3/2}}.
\end{eqnarray}
The surface integral can be written as,
\begin{eqnarray}
I_{\Upsilon}&=&\frac{1}{\Sigma}\oint_{\partial\Sigma}dl \frac{\partial f_{\Upsilon}}{\partial r}\nonumber\\
&=&\frac{1}{\Sigma}\int_{0}^{2\pi} R d\theta\left (\frac{\partial f_{\Upsilon}}{\partial r} \right)_{r=R}.
\end{eqnarray}
 Next we insert the derivative of the function $f_{\Upsilon}$ in the previous integral, and define the following parameters $A=(R^2+a_{0}^{2})R$, $B=-2a_{0}R^2$, $a=R^4+a_{0}^{2}R^2$ and $b=-2a_{0}R^3$.
 We consider the area $\Sigma=\pi R^2$ to obtain that,
\begin{eqnarray}
 I_{\Upsilon}=\frac{4}{R^2}.
\end{eqnarray}
The last integral to be calculated, $I_4$, defined in  Eq.(\ref{DefI4}), is the square of the superconductor density over the entire area of the cross section of the  cylinder.
Similarly to $I_2$, first we calculate the angular integral and define the results as a function of the radius, as shown in Eq.(\ref{Deff4r}).
To calculate the latter integral  we expand the numerator of the argument and divide the original integral in two parts, defined as follows: $f_4 (r) = f_{4}^{I}(r) + f_{4}^{II}(r)$, where the first part is given by,
\begin{eqnarray}\label{f4Ir}
 f_{4}^{I}(r)\equiv\int_{0}^{2\pi}d\theta\frac{(r^2+a_{0}^{2})^2-
 4\,a_0\,r(r^2+a_{0}^{2})\cos(\theta)}{\left(R^4+a_{0}^{2} \,r^2-2\,a_0\,r\,R^2\,\cos(\theta)\right)^2},\nonumber\\
\end{eqnarray}
and the second part is given by,
\begin{eqnarray}\label{f4IIr}
 f_{4}^{II}(r)\equiv\int_{0}^{2\pi}d\theta\frac{4\,a_{0}^{2} \,r^{2}\cos^{2}(\theta)}{\left(R^4+a_{0}^{2}\,r^2-2\,a_0\,r\,R^2\,\cos(\theta)\right)^2}. \nonumber\\
\end{eqnarray}
The integral (\ref{f4Ir}) is of the same type of the calculated integral (\ref{IntCos0_2pi}).
We just need to do $A=(r^2+a_{0}^{2})^2$, $B=-4a_0 r (r^2+a_{0}^{2})$, $a= R^4+a_{0}^{2}r^2$ and $b=-2a_{0} r R^2$.
The second integral we write as a derivative of an integral with a known result:
\begin{eqnarray}
 \int_{0}^{2\pi} dx \frac{B\cos^{2}(x)}{(a+b\cos(x))^2}=\frac{\partial}{\partial b}\int_{0}^{2\pi} dx \frac{-B\cos(x)}{a+b\cos(x)}.\nonumber\\
\end{eqnarray}
This integral is found in the right hand side of the previous equation and the result is given by,
\begin{eqnarray}
 \int_{0}^{2\pi} dx \frac{-B\cos(x)}{a+b\cos(x)}=2\pi\frac{B}{b}+2\pi\frac{a\;B}{b(a^2-b^2)},
\end{eqnarray}
where $A=0$.
Calculating the derivative, one finds the expression,
\begin{eqnarray}
 f_{4}^{II}=2\pi\frac{B}{b^2}-2\pi\frac{a\;B(a^2-2\;b^2)}{b^2(a^2-b^2)^{3/2}},
\end{eqnarray}
where the constants shown in the final result are related to the original expression for $f_{4}^{II}$ by $ B=4a_{0}^{2}r^{2} $, $ a=R^4 + a_{0}^{2}r^{2} $ and $ b=-2a_{0}r\;R^2$.
The final result for Eq.(\ref{Deff4r}) is a function of the radius and is given by
\begin{eqnarray}
 && f_{4}(r)=\frac{2\pi}{R^4}+\frac{2\pi}{R^4}\frac{1}{(R^4+a_{0}^{2}r^2)^3}\nonumber\\
 &&\times\Big[-\Big(1-\frac{R^4}{a_{0}^{4}}\Big)a_{0}^{6}r^6+\Big(7R^4 - 8\frac{R^6}{a_{0}^{2}}+\frac{R^8}{a_{0}^{4}}\Big)a_{0}^{4}r^4\nonumber\\
 &&+\Big(7R^8-8a_{0}^{2} R^6 + a_{0}^{4}R^4)a_{0}^{2}r^{2} - R^8 (R^4-a_{0}^{4})\Big].
\end{eqnarray}
To obtain $I_4$  we insert the function $f_4 (r)$ and we consider the following change of variable $ x=a_{0}^{2}r^{2}/R^{4}$.
With this change of variable and some simplifications one obtains that  $I_4=(R^2/a_{0}^{2})(a_{0}^{2}/R^2+ \sum_{j=0}^{3} d_j K_j)$, where the $d_j$ coefficients are defined as follows:
\begin{eqnarray}
 d_0&\equiv& -1 + \frac{a_{0}^{4}}{R^4},\\
 d_1&\equiv& 7 - 8 \frac{a_{0}^{2}}{R^2} + \frac{a_{0}^{4}}{R^4},\\
 d_2&\equiv& \frac{R^4}{a_{0}^{4}} - 8\frac{R^2}{a_{0}^{2}} + 7,\\
 d_3&\equiv& \frac{R^4}{a_{0}^{4}}-1,
\end{eqnarray}
and the integrals in the new variable $x$ are defined as
\begin{eqnarray}
 K_j\equiv \int_{0}^{a_{0}^{2}/R^2}\frac{x^{j} dx}{(1-x)^3}.
\end{eqnarray}
The calculated $K_j$ integrals give the results
\begin{eqnarray}
 K_0&=&\frac{1}{2}\left(1-\frac{a_{0}^{2}}{R^2}\right)^{-2}-\frac{1}{2},\\
 K_1&=&\frac{1}{2}\left(1-\frac{a_{0}^{2}}{R^2}\right)^{-2}-\left(1-\frac{a_{0}^{2}}{R^2}\right)^{-1}+\frac{1}{2},\\
 K_2&=&\frac{1}{2}\left(1-\frac{a_{0}^{2}}{R^2}\right)^{-2}-2\left(1-\frac{a_{0}^{2}}{R^2}\right)^{-1}+\frac{3}{2}\nonumber\\
 &-&\ln\left(1-\frac{a_{0}^{2}}{R^2}\right),\\
 K_3&=&\frac{1}{2}\left(1-\frac{a_{0}^{2}}{R^2}\right)^{-2}-3\left(1-\frac{a_{0}^{2}}{R^2}\right)^{-1}+\frac{5}{2}\nonumber\\
 &-&\frac{a_{0}^{2}}{R^2}-3\ln\left(1-\frac{a_{0}^{2}}{R^2}\right).
\end{eqnarray}
Then the products $ c_j K_j $ are obtained and summed over in $j$. After some simplifications  Eq.(\ref{I4final}) is finally obtained.
\bibliography{reference}
\end{document}